\documentclass[apj]{emulateapj}
\usepackage{multirow}

\shorttitle{Improved and Quality Checked Line measurements of SDSS galaxies}
\shortauthors{Oh et al.}

\begin{document}

\title{Improved and Quality-Assessed Emission and Absorption Line
  Measurements in Sloan Digital Sky Survey galaxies}

\author{Kyuseok Oh\altaffilmark{1}, Marc Sarzi\altaffilmark{2}, Kevin Schawinski\altaffilmark{3,4}, Sukyoung K. Yi\altaffilmark{1}}
\altaffiltext{1}{Department of Astronomy, Yonsei University, Seoul 120-749, Korea; yi@yonsei.ac.kr} 
\altaffiltext{2}{Centre for Astrophysics Research, University of Hertfordshire, Hatfield AL10 9AB, UK} 
\altaffiltext{3}{Yale Center for Astronomy and Astrophysics, Yale University, P.O. Box 208121, New Haven, CT 06511, U.S.A.}
\altaffiltext{4}{Einstein Fellow}

% Forbidden Lines
\def\OI{[\mbox{O\,{\sc i}}]~$\lambda 6300$}
\def\OIII{[\mbox{O\,{\sc iii}}]~$\lambda 5007$}
\def\OIIIs{[\mbox{O\,{\sc iii}}]~$\lambda 4363$}
\def\OIIIab{[\mbox{O\,{\sc iii}}]$\lambda\lambda 4959,5007$}
\def\SIIab{[\mbox{S\,{\sc ii}}]~$\lambda\lambda 6717,6731$}
\def\SII{[\mbox{S\,{\sc ii}}]~$\lambda \lambda 6717,6731$}
\def\NII{[\mbox{N\,{\sc ii}}]~$\lambda 6584$}
\def\NIIb{[\mbox{N\,{\sc ii}}]~$\lambda 6584$}
\def\NIIa{[\mbox{N\,{\sc ii}}]~$\lambda 6548$}
\def\NI{[\mbox{N\,{\sc i}}]~$\lambda \lambda 5198,5200$}

% for table 1 
\def\OIIa{[\mbox{O{\sc ii}}]~$\lambda 3726$}
\def\OIIb{[\mbox{O{\sc ii}}]~$\lambda 3729$}
\def\NeIIIa{[\mbox{Ne{\sc iii}}]~$\lambda 3869$}
\def\NeIIIb{[\mbox{Ne{\sc iii}}]~$\lambda 3967$}
\def\OIIIa{[\mbox{O{\sc iii}}]~$\lambda 4959$}
\def\OIIIb{[\mbox{O{\sc iii}}]~$\lambda 5007$}
\def\HeII{{He{\sc ii}}~$\lambda 4686$}
\def\ArIVa{[\mbox{Ar{\sc iv}}]~$\lambda 4711$}
\def\ArIVb{[\mbox{Ar{\sc iv}}]~$\lambda 4740$}
\def\NIa{[\mbox{N{\sc i}}]~$\lambda 5198$}
\def\NIb{[\mbox{N{\sc i}}]~$\lambda 5200$}
\def\HeI{{He{\sc i}}~$\lambda 5876$}
\def\OI{[\mbox{O{\sc i}}]~$\lambda 6300$}
\def\OIb{[\mbox{O{\sc i}}]~$\lambda 6364$}
\def\SIIa{[\mbox{S{\sc ii}}]~$\lambda 6716$}
\def\SIIb{[\mbox{S{\sc ii}}]~$\lambda 6731$}
\def\ArIII{[\mbox{Ar{\sc iii}}]~$\lambda 7136$}

% Balmer lines
\def\Ha{{H$\alpha$}}
\def\Hb{{H$\beta$}}

% Line ratios
\def\NIIHa{[\mbox{N\,{\sc ii}}]/H$\alpha$}
\def\SIIHa{[\mbox{S\,{\sc ii}}]/H$\alpha$}
\def\OIHa{[\mbox{O\,{\sc i}}]/H$\alpha$}
\def\OIIIHb{[\mbox{O\,{\sc iii}}]/H$\beta$}

% Common terms
\def\Ebmv{E($B-V$)}
\def\LOIII{$L[\mbox{O\,{\sc iii}}]$}
\def\Ledd{${L/L_{\rm Edd}}$}
\def\LOIIIs4{$L[\mbox{O\,{\sc iii}}]$/$\sigma^4$}
\def\LOIIIMbh{$L[\mbox{O\,{\sc iii}}]$/$M_{\rm BH}$}
\def\Mbh{$M_{\rm BH}$}
\def\Msigma{$M_{\rm BH} - \sigma$}
\def\Ms{$M_{\rm *}$}
\def\Msun{$M_{\odot}$}
\def\Msunyr{$M_{\odot}yr^{-1}$}

% Units
\def\ergs{$~\rm ergs^{-1}$}
\def\kms{${\rm km}~{\rm s}^{-1}$}
% Marc's definitions
%\newcommand{\kms}{\mbox{${\rm km\;s^{-1}}$}}
\newcommand{\cms}{\mbox{${\rm cm\;s^{-1}}$}}
\newcommand{\pccm}{\mbox{${\rm cm^{-3}}$}}
\newcommand{\sauron}{{\texttt {SAURON}}}
\newcommand{\oasis}{{\texttt {OASIS}}}
\newcommand{\HST}{{\it HST\/}}
\newcommand{\Vg}{$V_{\rm gas}$}
\newcommand{\Sg}{$\sigma_{\rm gas}$}
\newcommand{\eg}{e.g.,}
\newcommand{\ie}{i.e.,}

\newcommand{\gandalf}{{\texttt {gandalf}}}
\newcommand{\fracDeV}{{\texttt {FracDeV}}} 
\newcommand{\ppxf}{{\texttt {pPXF}}}

\begin{abstract}

% BEGIN Marc 
We present a new database of absorption and emission-line measurements
based on the entire spectral atlas from the Sloan Digital Sky Survey (SDSS) 
7$^{th}$ data release of galaxies within a redshift of 0.2.
Our work makes use of the publicly available penalized pixel-fitting
({\tt pPXF}) and gas and absorption line fitting ({\tt gandalf}) IDL
codes, aiming to improve the existing measurements for stellar
kinematics, the strength of various absorption-line features, and 
the flux and width of the emissions from different species of
ionised gas. 
Our fit to the stellar continuum uses both standard stellar population
models and empirical templates obtained by combining a large number
of stellar spectra in order to fit a subsample of high-quality SDSS
spectra for quiescent galaxies. Furthermore, our fit to the nebular
spectrum includes an exhaustive list of both recombination and
forbidden lines. Foreground Galactic extinction is implicitly treated
in our models, whereas reddening in the SDSS galaxies is included in
the form of a simple dust screen component affecting the entire
spectrum that is accompanied by a second reddening component affecting
only the ionised gas emission.
In order to check for systematic departures from the rather standard
set of assumptions that enters our models, we provide a quality assessment 
for our fit to the SDSS spectra in our sample, 
for both the stellar continuum and the nebular emissions and across different wavelength regions.
This quality assessment also allows the identification of objects with either
problematic data or peculiar features. 
We hope to foster the discovery potential of our database ; therefore, 
our spectral fit is available to the community.
For example, based on the quality assessment around the \Ha\ and
\NII\ lines, approximately 1\% of the SDSS spectra
classified as ``galaxies'' by the SDSS pipeline do in fact require
additional broad lines to be matched, even though they do not show
a strong continuum from an active nucleus, as do the SDSS objects
classified as ``quasars''.
Finally, we provide new spectral templates for galaxies
of different Hubble types, obtained by combining the results of our
spectral fit for a subsample of 452 morphologically selected objects.
% END Marc 
\end{abstract}	


%====================================================================
\section{Introduction}

% BEGIN Marc 
One of the most powerful ways to unveil the formation and evolution of
% END Marc
galaxies is to study their spectra. Since William Wollaston first
observed sunlight passing through a prism in 1802 to reveal a rich
spectrum, spectral lines have been used to understand the physical
processes governing the evolution of stars and galaxies. Galaxy
spectra feature stellar absorption and nebular emission lines, both of
which have been studied extensively over the past several decades.

% BEGIN Marc 
The prominent absorption features seen in galaxy spectra reflect the
average surface gravities, effective temperatures, metal abundances and
kinematics of their stellar components and thus hold a preserved record
of a galaxy's formation and evolution. In order to measure the
strength of various stellar absorption lines so as to compare them across
different objects and facilitate the prediction of theoretical models we
require a common system of integrated light spectroscopy, the most
prominent of which is the Lick/IDS system
\citep{bur84,gor93,wor94,wor97}. 
% END MARC
This widely-used system spans the optical wavelength regime from CN at
$\sim$4150\AA\ to TiO$_{2}$ at $\sim$6200\AA\ and defines a total of 25 standard absorption line indices. Measurements of these absorption
line strengths, combined with stellar population models have been
effectively used to study galaxy properties, 
such as : luminosity-weighted ages, metallicities and abundance ratios of
early-type galaxies \citep{dav93,fis95,fis96,jor99,kun00,kun06}.

% BEGIN Marc 
Nebular emission lines, on the other hand, probe the physical state of
the ionized gas in galaxies and thus can be used to trace 
the nuclear activity of, for example, a central supermassive black
hole or their instantaneous rate of star formation \citep{ost85}. 
Emission-line ratios have been turned into powerful
diagnostic tools, not just for individual galaxies, but for massive
spectroscopic surveys. \citet{bal81} first proposed the use of diagnostic diagrams, which have subsequently been used and refined by
numerous research groups \citep{vei87,kew01,kau03,sch07b}.
% END Marc 

Throughout the past two decades the Sloan Digital Sky Survey (SDSS) has
established the largest and most homogeneous database of both
photometric and spectroscopic data for galaxies. Using the Apache
Point Observatory's 2.5 m telescope, the SDSS has obtained 9400
$deg^{2}$ of imaging in the $u$, $g$, $r$, $i$ and $z$ bands
\citep{fuk96,gun98,gun06} on the AB magnitude system \citep{smi02}, considering both astrometric \citep{pie03} and photometric
\citep{ive04} calibrations. Follow-up spectroscopy
\citep{eis01,str02,ric02,sto02,bla03} has furthermore yielded nearly
one million objects spectroscopically classified as ``galaxies'' (See
Tab.~1 and Fig.~1 of \citealt{aba09}). The seventh, and latest, data
release of the SDSS (SDSS DR7, \citealt{aba09}) is notably improved, 
in terms of both photometry and spectroscopy, and is to date the largest 
sized database.

% BEGIN Marc
%
The SDSS spectroscopic database has been useful in numerous
investigations; however, the SDSS pipeline measurements still suffer from a
few crucial shortcomings. The most serious is that the
pipeline values for the absorption-line strength are not corrected for the
impact of nebular emissions. Not accounting for emission infills leads to an underestimation of the true strength of various absorption-line features and, in objects with strong emission lines - such as star-bursting systems or active nuclei - a more refined treatment of the nebular component may be needed to secure an unbiased extraction of the stellar kinematic. 
Furthermore, when measuring the ionised-gas emission it is
desirable to include reddening by dust and impose a sensible \textit{a priori}
on the relative strength of recombination lines, such as those from the
Balmer series.
The MPA-JHU group have addressed some of these issues
\citep[e.g.,][]{tre04}, refining the absorption-line strength
measurements for SDSS galaxies and leading to many useful studies
\citep[e.g.,][]{kau04,cid05,kew06,erb06,fab07}.
Yet, in the case of both these databases, it is not clear how
systematic problems may have affected the extraction of the physical
parameters listed in these catalogues, and it is not always possible
to verify the quality of these measurements themselves.
In this paper we attempt to further improve both the
absorption and emission-line spectroscopic measurements of nearby SDSS
galaxies, as well as providing an assessment of the quality of our fit
to both the stellar and nebular spectral components of our sample
galaxies. This helps to isolate objects with problematic data or
systems that cannot be described by our set of physical assumptions and
which may thus contain additional structures. For instance, in this paper 
we will show how such a quality assessment will reveal a previously unidentified population 
of objects with broad Balmer lines. Similar inspections may help finding the presence of other 
additional features such as the typical ``bump'' at 4640\AA\ associated to the presence of 
Wolf-Rayet stars (e.g., \citealt{bri08}) or a systematic mismatch in the NaD region 
around 5900\AA\ due to the presence of neutral material entrained in galactic outflows 
(e.g., \citealt{sat09}, \citealt{che10}). In fact, it is with the
hope of fostering the potential discovery of such peculiar systems
that we enable the community to inspect each of our fits with the SDSS galaxies considered in this study.

This work, unlike the other databases, adopt both theoretical stellar population synthesis model \citep{bru03} and empirical stellar templates \citep{san06} to improve accuracy of the fit. 
Besides, we treat nebular emission which affects all over the optical wavelengths 
including \NI\ doublet that is not considered by MPA-JHU group. 
Dramatic effect of \NI\ doublet subtraction on Mgb absorption line strength is described in \S~\ref{sec:abs_measurements}. 
Specifically, our sample includes the entire SDSS DR7
spectroscopic database of objects that the SDSS pipeline identified as
``galaxies'', which we restricted by imposing a redshift cut below $z
< 0.2$. This yielded a total of 661,400 galaxies.
%
% QUESTION: Any reason for this 0.2 cut? Why not 0.3, or 0.21 or no cut at all?}

This work is organized as follows. In \S~\ref{sec:method}, we describe
our method for deriving the absorption and emission line measurements,
which includes extracting the central stellar and gas kinematics.
In \S~\ref{sec:quality} we explain how we assessed the quality of our
fit in both the stellar continuum and the nebular emission, and explain how to identify 
and deal with objects with broad Balmer lines. The structure of our new database is then described in \S~\ref{sec:database}, with \S~\ref{sec:comparison} containing the details of how our measurements for the stellar kinematics, the emission-line fluxes and
widths, and the strength of absorption-line features compare with
the same quantities from the SDSS DR7 pipeline or the MPA-JHU DR7
catalogue. As an example of the improvements made by using our procedure, at the end of \S~\ref{sec:comparison} we discuss how accounting for the presence of the weak \NI\ emission can impact estimates of the element abundances in galaxies. Finally, we present a number of stacked spectra for different Hubble types and discuss how they can be useful to galaxy studies in \S~\ref{sec:stack} and give a summary of our work in \S~\ref{sec:summary}.

%\label{sec:method}
%\label{sec:quality}
%\label{ssec:contfit}
%\label{ssec:nebfit}
%\label{ssec:BLRfit}
%\label{sec:comparison}
%\label{sec:stack}
%\label{ssec:stack_sample}
%\label{ssec:stack_morpholoy}
% END Marc

%================================================================
\section{Method}
\label{sec:method}

In order to measure the stellar and gaseous kinematics within the central
regions of our sample galaxies and assess the depth of the most
important stellar absorption-line features, as well as the strength of
the nebular emission observed in the SDSS spectra, we adopted a strategy
very similar to that used during the \sauron\ integral-field
spectroscopic survey of nearby early-type galaxies
\citep{ems04,sar06,kun06}, which dealt with more than 34,000
individual spectra.

Our analysis of the SDSS spectra consisted of three separate steps.
First, we extracted the stellar kinematics by directly matching the
spectra with a set of stellar templates, while excluding the
spectral regions potentially affected by nebular emission.
We then measured the gaseous kinematics and assessed 
the strength of each emission-line by fitting the entire spectrum, 
simultaneously using the stellar templates that were broadened 
and shifted by the previously derived stellar kinematics and a set of 
Gaussian templates representing the nebular emission.
Finally, we subtracted, from the SDSS spectra, our best model for the
nebular emission and measured the strength of various stellar
absorption lines from the observed spectra following standard
line-strength index definitions.
An early application and description of this fitting strategy can be
found in \citet{sch07b} and \citet{car09}.

\subsection{Stellar Kinematics}

To measure the stellar kinematics from the SDSS spectra in our
catalogue, we used Cappellari \& Emsellem's (2004) pixel-fitting method, 
known as \ppxf, and parameterized the line-of-sight velocity
distribution (LOSVD) by means of a simple Gaussian, thus obtaining only
the galaxy redshift $z$ and central stellar velocity dispersion
$\sigma_{\star}$.
During the \ppxf\ fit, we used the \citet{bru03} stellar population models used by \citet{tre04}, which we combined with a number of empirical stellar templates based on the MILES stellar library \citep{san06}. 
The use of the MILES library helps with matching absorption-line features dominated by alpha-enhanced elements (see \S2.4).
Before extracting the central LOSVD in our sample galaxies, we also
needed to define the spectral regions used during the \ppxf\ fit.
This means setting the wavelength range of the fit, excluding the
spectral regions potentially affected by nebular emission, masking the
regions where the skylines at 5577\AA, 6300\AA, and 6363\AA (in the
rest frame) may have left substantial residuals after their
subtraction, and finally excluding the NaD$\lambda\lambda5890,5896$
absorption lines that were not properly matched by the
stellar templates owing to the impact of interstellar absorption.
The wavelength range of the \ppxf\ fit was set by the spectral coverage
of our templates, which is restricted to 6800\AA\ owing to the presence of the MILES templates, 
and the redshift estimate provided by the SDSS for each object, which also helped with the placement of a 1,200 \kms-wide mask (for emission-lines with a velocity dispersion of 200 \kms) centered around the expected position of each of the lines listed in Tab.~\ref{tab:emission}.
During the \ppxf\ fit, we used a 4$^{th}$-order additive polynomial
correction for the spectral slope of our templates.

%----------Table 1
\begin{center}
\begin{deluxetable}{cccl}
\tabletypesize{\scriptsize}
\tablecaption{Ionized-gas emission lines}
\tablewidth{0pt}
\tablehead{
\colhead{No} &
\colhead{Species} &
\colhead{Wavelength (\AA)} &
\colhead{relative strength to (line)}
}
\startdata
 1 &	 [\mbox{O{\sc ii}}]    & 3726.03 &       \\ 
 2 &	 [\mbox{O{\sc ii}}]    & 3728.73 &       \\
 3 &	 [\mbox{Ne{\sc iii}}]  & 3868.69 &       \\
 4 &	 [\mbox{Ne{\sc iii}}]  & 3967.40 &       \\
 5 &     H5   	               & 3889.05 & 0.037 (\Ha)\\
 6 &     H$\epsilon$           & 3970.07 & 0.056 (\Ha)\\
 7 &     H$\delta$             & 4101.73 & 0.091 (\Ha)\\
 8 &     H$\gamma$             & 4340.46 & 0.164 (\Ha)\\
 9 &     [\mbox{O{\sc iii}}]   & 4363.15 &       \\
10 &	 [\mbox{He{\sc ii}}]   & 4685.74 &       \\
11 &     [\mbox{Ar{\sc iv}}]   & 4711.30 &       \\
12 &     [\mbox{Ar{\sc iv}}]   & 4740.10 &       \\
13 &	 H$\beta$	       & 4861.32 & 0.350 (\Ha)\\
14 &	 [\mbox{O{\sc iii}}]   & 4958.83 & 0.350 (\OIII)\\
15 &	 [\mbox{O{\sc iii}}]   & 5006.77 &       \\
16 &     [\mbox{N{\sc i}}]     & 5197.90 &       \\
17 &     [\mbox{N{\sc i}}]     & 5200.39 &       \\
18 &	 [\mbox{He{\sc i}}]    & 5875.60 &       \\
19 &     [\mbox{O{\sc i}}]     & 6300.20 &       \\
20 &     [\mbox{O{\sc i}}]     & 6363.67 & 0.333 (\OI) \\
21 &	 [\mbox{N{\sc ii}}]    & 6547.96 & 0.340 (\NII)\\
22 &	 H$\alpha$ 	       & 6562.80 &       \\
23 &	 [\mbox{N{\sc ii}}]    & 6583.34 &       \\
24 &	 [\mbox{S{\sc ii}}]    & 6716.31 &       \\
25 & 	 [\mbox{S{\sc ii}}]    & 6730.68 &       \\
26 &	 [\mbox{Ar{\sc iii}}]  & 7135.67 &       \\
\enddata
\label{tab:emission}
\end{deluxetable}
\end{center}

%----------Table 2
\begin{deluxetable*}{crccl}
\tablewidth{0pt}
\tablecaption{Absorption line indices}
\tablehead{
\colhead{No} &
\colhead{Name} &
\colhead{Index Bandpass(\AA)} &
\colhead{Pseudocontinua(\AA)} &
\colhead{Reference} 
}
\startdata
01&$\rm{H\delta_{A}}$   &4083.500-4122.250	&4041.600-4079.750, 4128.500-4161.000	&Lick\tablenotemark{a}\\
02&$\rm{H\delta_{F}}$	&4091.000-4112.250	&4057.250-4088.500, 4114.750-4137.250	&Lick\\
03&$\rm{CN_{1}}$     	&4142.125-4177.125	&4080.125-4117.625, 4244.125-4284.125	&Lick\\
04&$\rm{CN_{2}}$     	&4142.125-4177.125	&4083.875-4096.375, 4244.125-4284.125	&Lick\\
05&Ca4227       	&4222.250-4234.750	&4211.000-4219.750, 4241.000-4251.000	&Lick\\
06&G4300        	&4281.375-4316.375	&4266.375-4282.625, 4318.875-4335.125	&Lick\\
07&$\rm{H\gamma_{A}}$	&4319.750-4363.500	&4283.500-4319.750, 4367.250-4419.750	&Lick\\
08&$\rm{H\gamma_{F}}$	&4331.250-4352.250	&4283.500-4319.750, 4354.750-4384.750	&Lick\\
09&Fe4383       	&4369.125-4420.375	&4359.125-4370.375, 4442.875-4455.375	&Lick\\
10&Ca4455       	&4452,125-4474.625	&4445.875-4454.625, 4477.125-4492.125	&Lick\\
11&Fe4531       	&4514.250-4559.250	&4504.250-4514.250, 4560.500-4579.250	&Lick\\
12&C4668        	&4634.000-4720.250	&4611.500-4630.250, 4742.750-4756.500	&Lick\\
13&$\rm{H\beta}$     	&4847.875-4876.625	&4827.875-4847.875, 4876.625-4891.625	&Lick\\
14&$\rm{H\beta_{G}}$ 	&4851.320-4871.320	&4815.000-4845.000, 4880.000-4930.000	&Gonzalez\tablenotemark{b}\\
15&Fe4930       	&4903.000-4945.500	&4894.500-4907.000, 4943.750-4954.500	&Gonzalez\\
16&$\rm{OIII_{1}}$   	&4948.920-4978.920	&4885.000-4935.000, 5030.000-5070.000	&Gonzalez\\
17&$\rm{OIII_{2}}$   	&4996.850-5016.850	&4885.000-4935.000, 5030.000-5070.000	&Gonzalez\\
18&Fe5015       	&4977.750-5054.000	&4946.500-4977.750, 5054.000-5065.250	&Lick\\
19&$\rm{Mg_{1}}$     	&5069.125-5134.125	&4895.125-4957.625, 5301.125-5366.125	&Lick\\
20&$\rm{Mg_{2}}$     	&5154.125-5196.625	&4895.125-4957.625, 5301.125-5366.125	&Lick\\
21&$\rm{Mg_{b}}$     	&5160.125-5192.625	&5142.625-5161.375, 5191.375-5206.375	&Lick\\
22&Fe5270       	&5245.650-5285.650	&5233.150-5248.150, 5285.650-5318.150	&Lick\\
23&$\rm{Fe5270_{S}}$  	&5256.500-5278.500	&5233.000-5250.000, 5285.500-5308.000	&SAURON\tablenotemark{c}\\
24&Fe5335       	&5312.125-5352.125	&5304.625-5315.875, 5353.375-5363.375	&Lick\\
25&Fe5406       	&5387.500-5415.000	&5376.250-5387.500, 5415.000-5425.000	&Lick\\
26&Fe5709       	&5696.625-5720.375	&5672.875-5696.625, 5722.875-5736.625	&Lick\\
27&Fe5782       	&5776.625-5796.625	&5765.375-5775.375, 5797.875-5811.625	&Lick\\
28&NaD	        	&5876.875-5909.375	&5860.625-5875.625, 5922.125-5948.125	&Lick\\
29&$\rm{TiO_{1}}$    	&5936.625-5994.125	&5816.625-5849.125, 6038.625-6103.625	&Lick\\
30&$\rm{TiO_{2}}$    	&6189.625-6272.125	&6066.625-6141.625, 6372.625-6415.125	&Lick\\
\enddata
\tablenotetext{a}{\citealt{wor94}}
\tablenotetext{b}{\citealt{gon93}}
\tablenotetext{c}{\citealt{kun06}}
\label{tab:absorption}
\end{deluxetable*}

\subsection{Emission Line Measurements}

Once the stellar kinematics had been measured, 
we lifted the emission-line mask and proceeded to simultaneously match 
the stellar continuum and the nebular emission in our sample galaxies
using Sarzi et al.'s (2006) \gandalf\ code.
During this matching process, \gandalf\ combines the stellar templates
with a number of emission-line templates represented either by single
Gaussians, in the case of single emission-line species, or by sets of
Gaussians meant to describe either doublets, such as \OIIIab, 
or multiplets, such as the entire Balmer series. The intrinsic relative
strength of the lines, prior to any reddening due to dust in the
target galaxies, was set by either atomic physics, in the case of
doublets, or by assuming a gas temperature, in the case of the Balmer
lines.
%
% BEGIN Marc
During the \gandalf\ fit the position and the intrinsic width (prior
to convolution with the line-spread function of the SDSS spectra) of
the Gaussian templates were solved non-linearly by means of a standard
Levenberg-Marquardt minimization\footnote{For such non-linear fit both
  \ppxf\ and \gandalf\ use the {\tt MPFIT} IDL routine provided by
  Craig Markwardt at {\tt http://cow.physics.wisc.edu/\~{}craigm/idl/}
}, while at each iteration of this process the relative strengths of
both the stellar and emission-line templates were solved linearly and by
imposing non-negative weights.
% END Marc
%
We imposed the same kinematics on all of the forbidden lines and the recombination lines. Only the gas kinematics was derived at
this stage. During the entire \gandalf\ fit the stellar templates
were shifted and broadened according to the previously derived stellar
LOSVD.
Reddening by dust was included in the \gandalf\ fit and was solved
non-linearly along with the gas kinematics. We included two
reddening components in our models: the first which affects the entire
spectrum, representing dust diffusion everywhere throughout the target
galaxy; and the second, which affects only the nebular emission and can account for dust more localized around the emission-line regions.
Finally, the NaD and skyline regions were excluded
during the \gandalf\ fit and, prior to that we de-reddened the SDSS
spectra to correct for foreground Galactic extinction (using maps from Schlegel et al. 1998) and to measure the intrinsic estimates
for dust reddening in our sample galaxies.
A complete list of the emission lines that were fit, including their
intrinsic relative strengths, is given in Tab.~\ref{tab:emission}.

\subsection{Absorption Line Measurements}

The last step of our analysis consisted of measuring the strength of
various stellar absorption lines throughout the wavelength range of
the SDSS spectra. This was done after subtracting our best \gandalf\ model for the
nebular emission and using the standard line-strength definition for
the indices listed in Tab.~\ref{tab:absorption}.
To perform the index measurements we used an IDL routine kindly
provided by H. Kuntschner.  Prior to the index calculation, this
routine brought the spectra back to the rest-frame and is capable of degrading 
their spectral resolution to match that of the Lick/IDS system, to
which most of the indices listed in Tab.~\ref{tab:absorption} belong.
Whether our line-strength measurements were based on the original
emission-line cleaned spectra or on their degraded counterparts (in
our catalogue we provide both values), we corrected our index
values for the impact of stellar kinematic broadening.
To achieve this effect, we observed how the values of the indices
changed when we measured them on the optimal combination of the stellar
templates returned by the \gandalf\ fit ($LS_{optimal}$) and on the best stellar model 
obtained during the same process ($LS_{model}$), which is the optimal template convolved by
the LOSVD obtained during the \ppxf\ fit and further adjusted for dust
reddening by \gandalf. Given the index measurement on the
emission-cleaned SDSS spectra ($LS$), the corresponding index value
corrected for kinematic broadening can be derived as follows

\begin{equation}
LS_{corr} = LS \times \frac{LS_{optimal}}{LS_{model}}
\end{equation}

Prior to the line-strength measurements we subtracted all of the emission 
fitted by \gandalf, even that from weak lines that would
be too weak to be formally deemed detected.  Although this may
produce an over-subtraction of the nebular flux and result in somewhat
biased values of the absorption-line indices, we prefer this to
introducing a sudden step in the index infill correction by not
subtracting the emission when crossing a certain detection threshold.
Moreover, when the skylines fall within the pseudocontinuum
or index passbands of one of our indices, which can happen for
different indices at different redshifts, we provided the $LS_{optimal}$
value that is measured on the stellar optimal template.  The residuals
of an imperfect subtraction of such skylines may lead to
meaningless values for the indices if these were measured directly on
the SDSS data.

\subsection{Optimization of Template Library}

As briefly mentioned at the beginning of this section, the templates
based on the stellar population models from Bruzual \& Charlot(2003)
cannot fully describe the stellar spectrum of our sample galaxies.
This is particularly true in cases of the most massive or the closest
objects in our sample, where the SDSS spectra probe overall or central
stellar populations that display an overabundance of alpha-elements,
which are not currently accounted for by the population models. Typical
spectral regions that are not well matched in these cases are those
around the CN 4200\AA\ band or the Mgb index.
If stellar population models cannot fully match such alpha-enhanced
galactic spectra, the combination of a large number of the stellar
spectra on which the models are based can.  Such stellar spectra
cannot be used individually together with the population models, 
as this would make the \ppxf\ and \gandalf\ fit excessively cumbersome. 
Furthermore, the resulting description of the
stellar continuum would not necessarily be physically motivated, which
in turn could lead to biased emission-line results and a wrong infill
correction for the absorption-line indices.
On the other hand, following \citet{sar10} we could select a number of
SDSS spectra of objects totally devoid of nebular emission to
construct, with the stellar library, a more limited number of
semi-empirical alpha-enhanced templates.
More specifically, using an initial subsample of our catalogue, 
we drew a number of passive galaxies covering a range of values for
the stellar velocity dispersion, to trace the galaxy mass, and for the
ratio of the Mgb and Fe5015 absorption-line indices, to trace the
degree of alpha-enhancement (see, e.g., \citealt{tho05}).  Using
\ppxf\ we then matched each of these spectra with the entire
MILES stellar library from \citet{san06}, without masking the
emission-line regions and while employing only multiplicative
polynomials to adjust the shape of the stellar templates. The quality
of such a fit is generally excellent, thanks to the large number
of templates, to the extent that the resulting optimal combinations of the
stellar spectra used in each of these fits could, in practice, be regarded
as a galactic spectrum deconvolved from the line-of-sight kinematical
broadening. Such optimal templates were then themselves combined into
12 empirical templates and designed to cover, as evenly as possible, the
distribution of the fitted galaxies in the $\sigma_{\star}$-Mgb/Fe5015
space.  Finally, these templates were convolved and re-binned in order
to match the spectral resolution and sampling in the \citet{bru03}
model templates.

%----------Figure 1
\begin{figure*}
\centering
\includegraphics[width=1\textwidth]{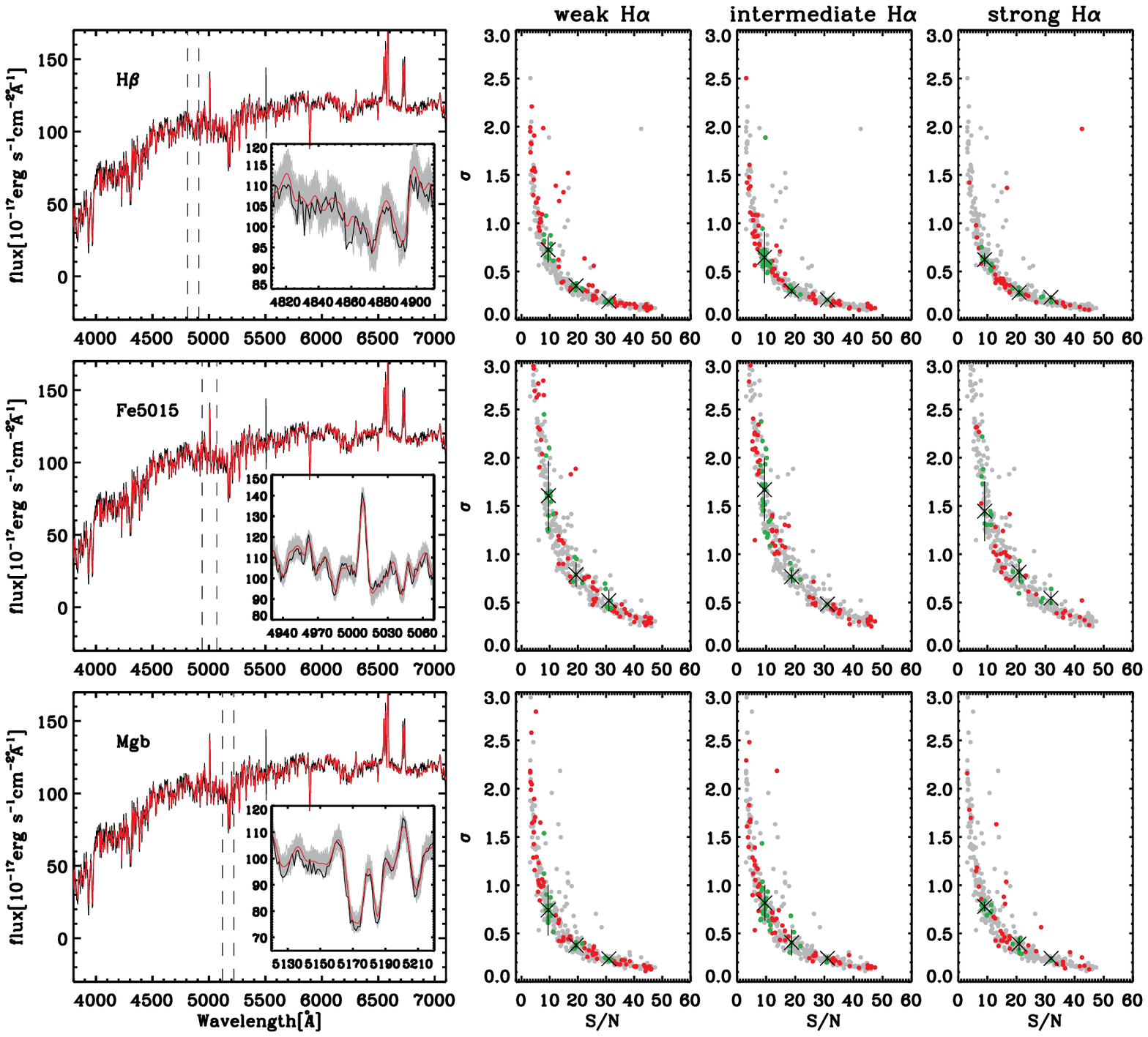}
\caption{Error estimation for absorption-line indices. The observed
  spectrum of a sample galaxy is shown in the left three panels,
  together with its best \gandalf\ fit (red lines). 
  In each of these panels, the inset shows the SDSS spectrum and our fit in a specific
  index region (for \Hb, Fe5015 and Mgb from top to bottom,
  respectively, and as indicated by the dashed vertical lines in each main panel), 
  as well as the flux density range of 100
  synthetic realizations of this particular spectrum (grey bands) based
  on the formal SDSS uncertainties on the flux density and our
  \gandalf\ fit.
 The right panels show the range (from top to bottom for the same indices as the left) 
 when considering a sample of 405 objects (grey points) with varying quality of spectra, 
 as traced by S/sN, and different strengths of nebular emission, as traced by the \Ha\ line. 
 From left to right, the red and green points highlight objects with increasing strength of \Ha\ emission, 
 with green points showing objects within specific S/sN ranges, where the median value of the errors on the absorption-line indices and the standard deviations are shown (crosses and vertical lines).}
\label{error}
\end{figure*}

\subsection{Errors}

Both the \ppxf\ and \gandalf\ returned formal errors for the stellar and
gaseous kinematics and the flux values of the fitted emission
lines. From MonteCarlo simulations based on synthetic spectra created
from our best \gandalf\ fit to a subsample of galaxies in our
catalogue and the formal uncertainties of their SDSS spectra, 
we verified that the formal uncertainties were consistent with the scatter
of the parameter estimations obtained from fitting the synthetic spectra.
Errors on the line-strength indices for each single object in our
database would be more cumbersome to compute. In fact, although the
impact of the statistic uncertainties on the flux densities of the
SDSS spectra could be easily described in terms of formal errors in
the line indices, the knock-on effect of the uncertainties on the
stellar kinematics and - most importantly - on the
\gandalf\ emission-line fit would require a MonteCarlo
simulation in order to be properly estimated.
Since this is not viable for nearly 700,000 objects, 
we decided instead to attach to any given line-index measurement in the
objects of our catalogue the typical uncertainty that is measured in
spectra showing a similar quality, a similar stellar kinematic
broadening and strength of the emission falling in the
pseudo-continuum of index passband of the considered index.
To derive such typical errors, we selected a large number of SDSS
spectra covering a wide range in quality (quantified by the average
value of the S/N ratio between the values of the flux density and the
statistical SDSS errors on them), of values of the stellar velocity
dispersion $\sigma_{\star}$, and - within similar S/N and
$\sigma_{\star}$ values - of strength for the emission possibly
affecting the line indices (e.g., \OIIIb\ for Fe5015). The strength of
the emission lines was quantified by the ratio A/rN of the best
fitting amplitude A of the considered line and the level of scatter rN
in the residuals of the \gandalf\ fit.
Based on our best \gandalf\ fit for these objects and the formal SDSS
errors on the flux densities, we created a number of synthetic
spectra with a slightly different kinematic broadening and nebular
emission, as allowed by the formal uncertainties on the \ppxf\ and
\gandalf\ parameters, and re-ran our entire analysis on them to obtain
new values for the line-indices.
% MARC
By re-matching each of these synthetic spectra and subsequently measuring the 
line-strength index after subtracting our best model for the nebular emission, 
we obtained a range of absorption-line strength values and the corresponding 
1$\sigma$-errors. 
The scatter of such synthetic line-index values around the one
measured on the original spectra provided us with the typical errors on the
line-strength measured in our catalogue. 
Fig.~\ref{error} shows how
their values vary as a function of S/N, $\sigma_{\star}$ and - when
applicable, the A/N of the contaminating lines. Furthermore, despite the presence of 
nebular emission, the error budget in the absorption-line indices is chiefly dominated by the 
quality of the spectra themselves. 
The formal uncertainties are also provided for the values of both
reddening components included in our \gandalf\ fit.
The typical errors for absorption-line indices are listed in Tab.~\ref{tab:absorption_error}.

%----------Table 3
\begin{turnpage}
\begin{table*}
\begin{center}
\caption{Typical errors for absorption-line indices}
\begin{tabular}{cccccccccc}
\tableline
\tableline
\multicolumn{1}{c}{}	&	\multicolumn{3}{c}{S/N\tablenotemark{a}$\sim10$}	&	\multicolumn{3}{c}{S/N$\sim20$}	&	\multicolumn{3}{c}{S/N$\sim30$}\\
\multicolumn{1}{c}{Name} & A/N\tablenotemark{b}$<$4		&	10$<$A/N$<$25	&	30$<$A/N$<$50	& A/N$<$4		&	10$<$A/N$<$25	&	30$<$A/N$<$50	&A/N$<$4		&	10$<$A/N$<$25	&	30$<$A/N$<$50	\\
\tableline
$\rm{H\delta_{A}}$   	&	1.434$\pm0.405$	&	1.046$\pm0.212$	&	0.859$\pm0.139$	&	0.462$\pm0.193$	&	0.574$\pm0.100$	&	0.439$\pm0.017$	&	0.323$\pm0.088$	&	0.332$\pm0.018$	&	0.375$\pm0.036$	\\
$\rm{H\delta_{F}}$    &	0.960$\pm0.294$	&	0.700$\pm0.158$	&	0.557$\pm0.087$	&	0.400$\pm0.066$	&	0.379$\pm0.051$	&	0.292$\pm0.040$	&	0.247$\pm0.055$	&	0.235$\pm0.018$	&	0.214$\pm0.023$	\\
$\rm{CN_{1}}$		&	0.038$\pm0.010$	&	0.031$\pm0.007$	&	0.028$\pm0.005$	&	0.013$\pm0.001$	&	0.018$\pm0.003$	&	0.013$\pm0.001$	&	0.009$\pm0.003$	&	0.010$\pm0.001$	&	0.010$\pm0.001$	\\
$\rm{CN_{2}}$	    	&	0.051$\pm0.013$	&	0.035$\pm0.007$	&	0.030$\pm0.005$	&	0.024$\pm0.006$	&	0.019$\pm0.003$	&	0.015$\pm0.001$	&	0.014$\pm0.003$	&	0.010$\pm0.000$	&	0.012$\pm0.001$	\\
Ca4227			&	0.540$\pm0.149$	&	0.442$\pm0.091$	&	0.382$\pm0.042$	&	0.259$\pm0.049$	&	0.258$\pm0.046$	&	0.193$\pm0.012$	&	0.153$\pm0.022$	&	0.168$\pm0.030$	&	0.152$\pm0.016$	\\
G4300		         &	1.180$\pm0.309$   	&	0.905$\pm0.159$	&	0.832$\pm0.125$	&	0.507$\pm0.073$	&	0.494$\pm0.067$	&	0.414$\pm0.065$	&	0.283$\pm0.027$	&	0.305$\pm0.028$ 	&	0.297$\pm0.030$	\\
$\rm{H\gamma_{A}}$&	1.345$\pm0.395$	&	1.019$\pm0.221$	&	0.899$\pm0.116$	&	0.568$\pm0.082$	&	0.559$\pm0.086$	&	0.468$\pm0.067$	&	0.352$\pm0.045$	&	0.338$\pm0.040$	&	0.435$\pm0.051$	\\
$\rm{H\gamma_{F}}$	&	0.866$\pm0.283$   	&	0.592$\pm0.114$   	&	0.538$\pm0.087$	&	0.356$\pm0.045$	&	0.330$\pm0.045$	&	0.264$\pm0.032$	&	0.214$\pm0.031$	&	0.184$\pm0.009$	&	0.225$\pm0.020$	\\
Fe4383       		&	1.454$\pm0.324$	&	1.242$\pm0.220$	&	1.160$\pm0.129$	&	0.599$\pm0.060$	&	0.621$\pm0.107$	&	0.571$\pm0.068$	&	0.417$\pm0.044$	&	0.387$\pm0.014$	&	0.444$\pm0.046$	\\
Ca4455       		&	0.655$\pm0.183$	&	0.559$\pm0.088$	&	0.523$\pm0.066$	&	0.315$\pm0.035$	&	0.312$\pm0.037$	&	0.257$\pm0.024$	&	0.181$\pm0.018$	&	0.205$\pm0.026$	&	0.211$\pm0.028$	\\
Fe4531       		&	1.048$\pm0.218$	&	1.020$\pm0.176$	&	0.968$\pm0.103$	&	0.493$\pm0.064$	&	0.511$\pm0.081$	&	0.426$\pm0.042$	&	0.316$\pm0.025$	&	0.313$\pm0.003$	&	0.328$\pm0.056$	\\
C4668        		&	1.912$\pm0.295$	&	1.745$\pm0.356$	&	1.742$\pm0.234$	&	0.848$\pm0.068$	&	0.881$\pm0.118$	&	0.758$\pm0.054$	&	0.493$\pm0.048$	&	0.564$\pm0.047$	&	0.578$\pm0.022$	\\
$\rm{H\beta}$     	&	0.726$\pm0.133$	&	0.646$\pm0.266$	&	0.623$\pm0.068$	&	0.352$\pm0.027$	&	0.299$\pm0.031$	&	0.283$\pm0.029$	&	0.190$\pm0.009$	&	0.208$\pm0.004$	&	0.232$\pm0.026$	\\
$\rm{H\beta_{G}}$ 	&	0.523$\pm0.110$	&	0.460$\pm0.180$	&	0.462$\pm0.049$	&	0.242$\pm0.027$	&	0.238$\pm0.037$	&	0.213$\pm0.021$	&	0.148$\pm0.009$	&	0.140$\pm0.004$	&	0.161$\pm0.012$	\\
Fe4930       		&	0.912$\pm0.384$	&	0.885$\pm0.237$	&	0.908$\pm0.105$	&	0.539$\pm0.100$	&	0.534$\pm0.231$	&	0.401$\pm0.049$	&	0.274$\pm0.025$	&	0.315$\pm0.013$	&	0.298$\pm0.012$	\\
$\rm{OIII_{1}}$   	&	0.693$\pm0.147$	&	0.743$\pm0.123$	&	0.649$\pm0.127$	&	0.350$\pm0.075$	&	0.402$\pm0.051$	&	0.309$\pm0.109$	&	0.205$\pm0.030$	&	0.244$\pm0.016$	&	0.223$\pm0.010$	\\
$\rm{OIII_{2}}$   	&	0.525$\pm0.114$	&	0.604$\pm0.296$	&	0.508$\pm0.300$	&	0.384$\pm0.101$	&	0.281$\pm0.086$	&	0.223$\pm0.218$	&	0.226$\pm0.128$	&	0.278$\pm0.100$	&	0.141$\pm0.043$	\\
Fe5015       		&	1.606$\pm0.361$	&	1.668$\pm0.327$	&	1.446$\pm0.310$	&	0.789$\pm0.127$	&	0.767$\pm0.070$	&	0.814$\pm0.105$	&	0.520$\pm0.111$	&	0.483$\pm0.009$	&	0.544$\pm0.066$	\\
$\rm{Mg_{1}}$     	&	0.021$\pm0.012$	&	0.020$\pm0.003$	&	0.020$\pm0.005$	&	0.009$\pm0.003$	&	0.010$\pm0.001$	&	0.009$\pm0.003$	&	0.006$\pm0.001$	&	0.006$\pm0.000$	&	0.007$\pm0.001$	\\
$\rm{Mg_{2}}$     	&	0.022$\pm0.006$	&	0.022$\pm0.006$	&	0.024$\pm0.006$	&	0.012$\pm0.002$	&	0.015$\pm0.004$	&	0.010$\pm0.002$	&	0.008$\pm0.001$	&	0.007$\pm0.001$	&	0.007$\pm0.001$	\\
$\rm{Mg_{b}}$     	&	0.742$\pm0.265$	&	0.816$\pm0.186$	&	0.779$\pm0.048$	&	0.377$\pm0.027$	&	0.406$\pm0.137$	&	0.391$\pm0.067$	&	0.237$\pm0.013$	&	0.243$\pm0.031$	&	0.242$\pm0.015$	\\
Fe5270       		&	0.931$\pm0.204$	&	0.904$\pm0.367$	&	0.910$\pm0.141$	&	0.405$\pm0.072$	&	0.447$\pm0.072$	&	0.480$\pm0.125$	&	0.260$\pm0.050$	&	0.275$\pm0.014$	&	0.304$\pm0.018$	\\
$\rm{Fe5270_{S}}$  	&	0.634$\pm0.154$	&	0.644$\pm0.291$	&	0.665$\pm0.082$	&	0.277$\pm0.048$	&	0.277$\pm0.043$	&	0.318$\pm0.124$	&	0.181$\pm0.038$	&	0.294$\pm0.158$	&	0.204$\pm0.010$	\\
Fe5335       		&	0.882$\pm0.288$	&	0.905$\pm0.149$	&	0.817$\pm0.208$	&	0.506$\pm0.068$	&	0.461$\pm0.048$	&	0.390$\pm0.050$	&	0.284$\pm0.050$	&	0.406$\pm0.080$	&	0.318$\pm0.038$	\\
Fe5406       		&	0.669$\pm0.252$	&	0.664$\pm0.299$	&	0.734$\pm0.094$	&	0.339$\pm0.072$	&	0.376$\pm0.075$	&	0.284$\pm0.078$	&	0.191$\pm0.023$	&	0.243$\pm0.004$	&	0.267$\pm0.089$	\\
Fe5709       		&	0.553$\pm0.152$	&	0.484$\pm0.144$	&	0.539$\pm0.109$	&	0.237$\pm0.013$	&	0.234$\pm0.037$	&	0.216$\pm0.035$	&	0.154$\pm0.030$	&	0.167$\pm0.021$	&	0.183$\pm0.018$	\\
Fe5782       		&	0.476$\pm0.102$	&	0.395$\pm0.071$	&	0.463$\pm0.087$	&	0.196$\pm0.080$	&	0.235$\pm0.030$	&	0.175$\pm0.035$	&	0.139$\pm0.006$	&	0.146$\pm0.010$	&	0.131$\pm0.023$	\\
NaD	        			&	0.601$\pm0.172$	&	0.530$\pm0.141$	&	0.644$\pm0.112$	&	0.271$\pm0.037$	&	0.353$\pm0.083$	&	0.260$\pm0.037$	&	0.175$\pm0.028$	&	0.225$\pm0.013$	&	0.191$\pm0.032$	\\
$\rm{TiO_{1}}$    	&	0.015$\pm0.004$	&	0.015$\pm0.004$	&	0.019$\pm0.005$	&	0.008$\pm0.001$	&	0.008$\pm0.001$	&	0.007$\pm0.001$	&	0.005$\pm0.001$	&	0.005$\pm0.000$	&	0.005$\pm0.001$	\\
$\rm{TiO_{2}}$    	&	0.013$\pm0.001$	&	0.013$\pm0.002$	&	0.015$\pm0.003$	&	0.006$\pm0.001$	&	0.007$\pm0.001$	&	0.006$\pm0.001$	&	0.004$\pm0.001$	&	0.004$\pm0.000$	&	0.005$\pm0.000$	\\
\tableline
\tablenotetext{a}{mean S/N of SDSS g, r and i-band}
\tablenotetext{b}{the best fitting amplitude A of the considered line and the level of scatter rN in the residuals of the \gandalf\ fit on $\rm{H\alpha}$}
\end{tabular}
\end{center}
\label{tab:absorption_error}
\end{table*}
\end{turnpage}

%\clearpage

\subsection{Final Remarks}

We conclude this section with a couple of remarks. 
First, before the strength of the absorption features were assessed at the end
of our procedure, the first two steps of our analysis were repeated in
order to optimize our \gandalf\ fit. In particular, once the position
and the width of the lines were measured the first time, we repeated
the \ppxf\ fit by placing a better emission-line mask on only the
region where some emission was found. This makes us more confident 
in the stellar kinematics, in particular for galaxies with little
or no emission or in star-forming objects where the Balmer absorption
lines are the predominant feature in the stellar spectrum and cannot
be properly matched unless the mask around the Balmer emission lines -
generally narrower than the absorption lines - is properly adjusted. 
A better stellar kinematic leads to a better \gandalf\ fit.

A second point to keep in mind is that the second component for
the dust extinction used in the \gandalf\ fit - that affecting only
the nebular spectrum - should only be considered reliable in the
presence of Balmer lines, and thus of an observed Balmer decrement.
If no Balmer lines are detected, or if only \Ha\ is, then the
de-reddened values for the fluxes of all the lines fitted by
\gandalf\ will also be very uncertain, and one should only use the
observed fluxes.
Finally, we stress that even in the presence of a Balmer decrement in
a good quality spectrum, the values of the diffuse component for the
reddening - approximated as a uniform screen affecting the entire
galaxy spectrum - may to some extent be degenerate with the particular
\gandalf\ choice for the stellar templates.
Determining the star-formation history and the exact values of the
reddening in our sample galaxies is beyond the scope of this catalogue,
however, and here we are interested mostly in providing the best
physically motivated and accurate description of the stellar and
nebular spectrum, in order to provide reliable values for the stellar
and gas kinematics, for the emission-line fluxes and for the
infill-corrected absorption-line indices.

%----------Figure 2
\begin{figure*}
\centering
\includegraphics[width=1\textwidth]{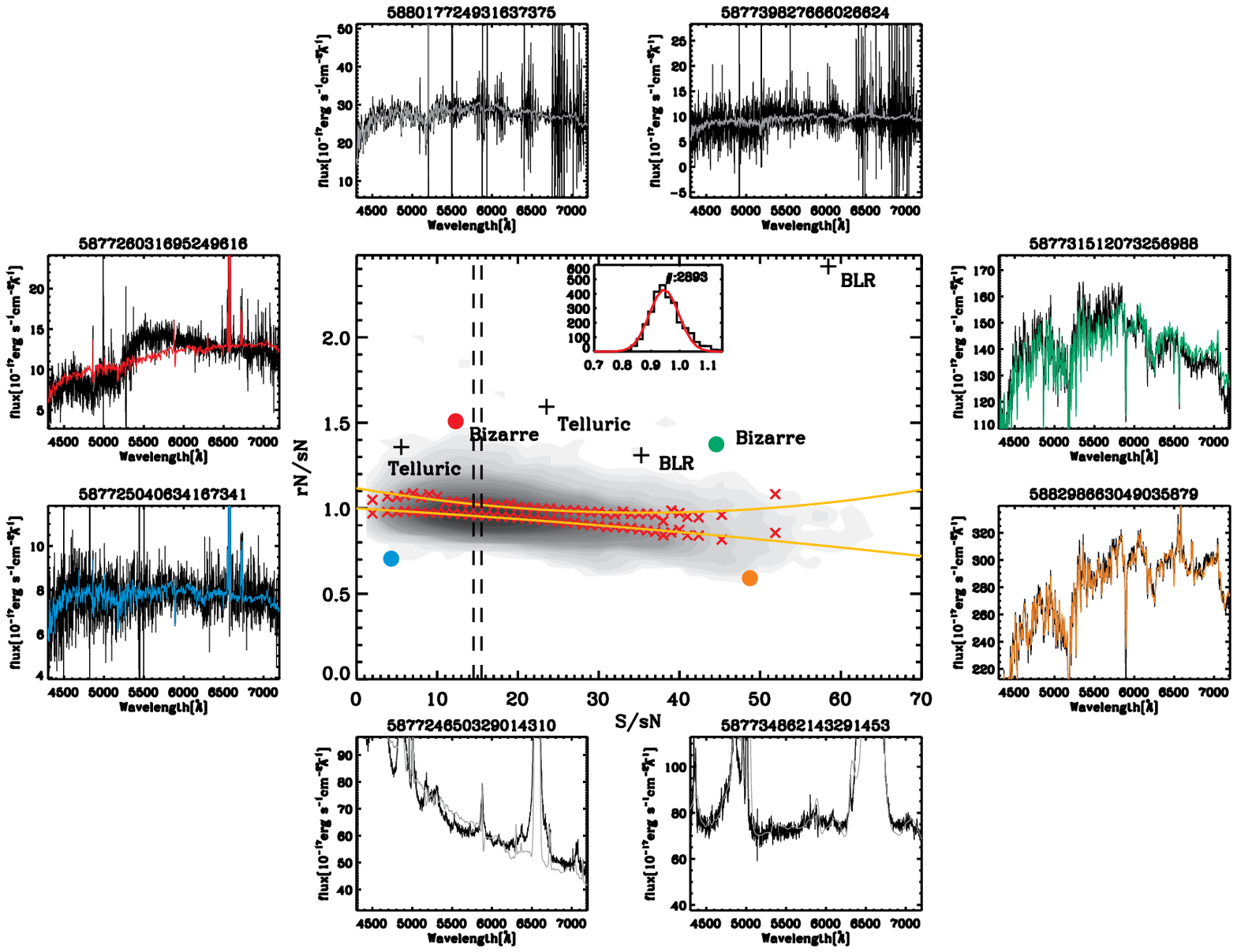}
\caption{Quality assessment for continuum with examples inducing bad
  fits. This figure shows a quality assessment for a continuum region 
  using randomly selected $\sim$46000 objects. Red
  crosses on the central panel denote the median (lower) and 1 $\sigma$ (upper)
  distribution for each S/sN bin and orange solid lines fit (note
  that there is a demarcation line for 1 $\sigma$ in the direction of
  larger rN/sN.). The vertical dashed lines indicate a specific bin and its
  Gaussian distribution has been inserted on the top side as an example. 
  The color filled dots correspond to the colored left and right panels and clarify 
  the trend for quality assessments. 
  The black solid lines represent the observed spectra and the coloured lines are the fits. 
  Furthermore, the top and bottom examples with gray fits are the major
  reasons for the bad fits. These telluric contaminated spectra (top) and
  Broad Line Regions (bottom) are marked with black crosses on the central
  panel. A minor reason which brought about bad fits is also denoted by 
  red and green filled dots.}
\label{qa_continuum}
\end{figure*}

%==============================================================
\section{Fitting quality assessment}
\label{sec:quality}

Assessing the quality of our fits to the SDSS spectra is a key to
the accuracy of our measurements for the stellar and gaseous
kinematics, as well as for the strengths of both the emission and
absorption lines. An inadequate model for the observed spectrum, or
artificial features in the data themselves is likely to
introduce biases to the measured parameters that are not accounted for 
by our formal errors.
Such parameters are affected in different ways by data mismatches. 
For the stellar kinematics, and in particular for the velocity
dispersion, the most crucial match is the stellar continuum
across the entire wavelength range, whereas the accuracy of the
emission-line parameters (and of the absorption-line indices that
are affected by nebular emission) are more sensitive to the quality of
the fit in more localized spectral regions around the emission
lines.

The quality of the data, as routinely estimated by dividing the
average level of the flux density (hereafter S, for signal) by the
level of the formal uncertainties for the latter (hereafter sN, for
statistical noise), does not guarantee a good fit.
Artifacts introduced during data reduction may not be
picked up by such an S/sN ratio, whereas a severe template mismatch to
the stellar continuum or the presence of additional components in the
nebular spectrum (e.g., a Broad Line Region or Wolf-Rayet features
that are not included in our standard fitting procedure) may lead to
biased results, even in a case of excellent data.
A more direct way to assess the quality of our model would be 
to compare the level of fluctuations in the fit residuals 
(hereafter rN, for residual noise) to the expected statistical fluctuations, sN. 
A rN/sN ratio close to unity indicates a good fit, as this
ration corresponds to a reduced $\rm{\chi^{2}}$, which is also close to 1.

%\placefigtwo

\subsection{Continuum fit}
\label{ssec:contfit}

In order to assess the quality of the fit to the stellar continuum and
help in isolating biased measurements for the stellar kinematics, we
measured the rN/sN ratio in four different wavelength regions of equal
width that were devoid of nebular emission. Specifically, we
adopted the 4500 -- 4700\AA, 5400 -- 5600\AA, 6000 -- 6200\AA\ and
6800 -- 7000\AA\ wavelength intervals and combined the corresponding
values for the rN/sN ratio into a single average, 
excluding the largest value in order to avoid a spurious contribution of
continuum bands, where the SDSS spectra happened to be broken.

%----------Figure 3
\begin{figure*}
\centering
\includegraphics[width=1\textwidth]{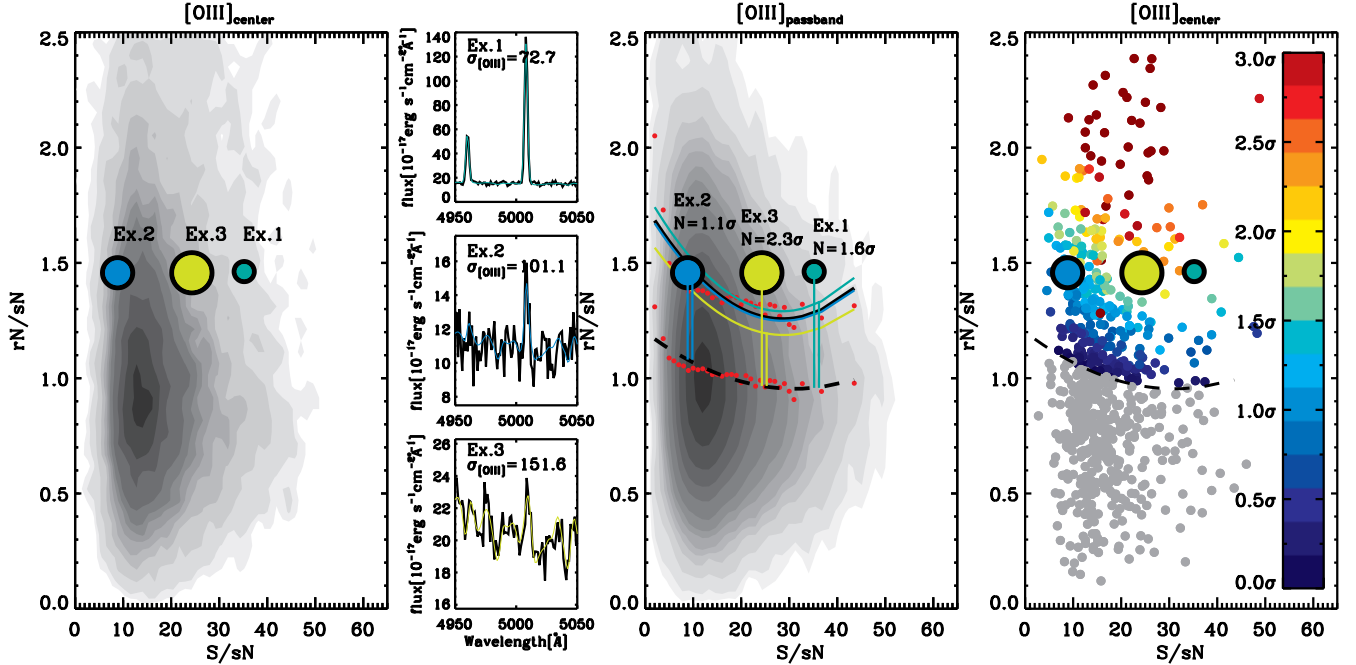}
\caption{Quality assessment process for \OIII\ with three examples.
  Left : The quality assessment plane drawn by a central emission region
  using randomly selected $\sim$46000 objects. The three examples which
  have different emission line widths are shown as Ex.1,2 and 3. Middle :
  Quality assessment plane given by a typical passband. 
  The lower and upper red filled dots are the median and 1$\sigma$ at each
  S/sN bin. The black dashed and solid lines trace each point.
  Moreover, \OIII\ emissions for these three examples are
  shown on left sub-panels in corresponding colors. The black
  line represents the observed spectrum and the colored one the fit. Once we
  derived the median and 1$\sigma$ from the typical passband, we measured the 
  N$\sigma$ of every \OIII\ emission. The newly derived 1$\sigma$ are also
  shown using the same colors. Right : 1,000 objects shown in different colors 
  depend on N$\sigma$. The objects below the median are indicated in gray.}
\label{qa_emission}
\end{figure*}

As shown in Fig.~\ref{qa_continuum}, for $\sim$46000 randomly selected SDSS
galaxy spectra, the distribution of the rN/sN average values for the
fit to the stellar continuum as a function of the quality of the
spectrum itself, as measured by the S/sN ratio, measured and averaged
in the same continuum bands. The vast majority of the spectra were well
matched by our model, with rN/sN close to unity. Our ability to match
the data also appears to have improved as the quality of the spectra
increased. In fact, for data of excellent quality (high S/sN), 
the level of fluctuation in the fit residual became even smaller than the
formal uncertainties in the flux density, suggesting that the latter
errors must under-predict the real level of statistical fluctuations
in the SDSS spectra in such high-quality regimes.

Fig.~\ref{qa_continuum} also shows a few representative examples of
\gandalf\ fits to the the SDSS spectra. As stated above, the quality of
the data themselves does not guarantee success or failure in the
fit.  It is possible to have formally good fits, both in cases of
data with good and relatively poor values for the S/sN ratio (orange
and blue examples), whereas our model did not necessarily fail only
in a case of poor data (red example), but also when the quality was
very good (green example).

In general there are three reasons that our objects show rN/sN ratios
that are significantly above unity.  Most often a poor fit to the
continuum is the result of the presence of a conspicuous Broad line
region (grey examples, lower panels), which was not noticed by the
SDSS pipeline in the process of separating the galaxy from the QSO
spectra. Although the presence of broad line regions (BLR) can be
accounted for by \gandalf\ (see \S~3.3), in the framework of our
standard recipe that includes only narrow-line regions, the presence of
an additional component leads to a considerable mismatch outside 
the wavelength regions near the \Ha\ and \Hb\ lines. A second
kind of failure occurs for objects in which the adopted stellar
library appears to be inadequate in matching the observed SDSS
spectrum (dubbed as ``Bizarre'' in Fig.~\ref{qa_continuum} and already
shown in the red and green examples). This occurs in a very small
fraction of objects ; however, it often appears the spectra have been 
contaminated by features in the sky foreground. 
Finally, for a third kind of spectra, a high rN/sN value
is the result of telluric atmospheric features at the red end of the
spectra. When such features are present, the SDSS formal errors underestimate 
the fluctuations around the stellar continuum
introduced by the telluric features, biasing the rN/sN value in the
6800-7000 continuum band to high values. We note however, that in this
case the fit to the rest of the spectrum is not necessarily poor.

Having ascertained that the vast majority of our models showed good fits for the
stellar continuum of the SDSS spectra, we could assign, to any object
with rN/sN above unity, a probability of being a true outlier (in the
distribution shown in Fig.~\ref{qa_continuum}), and hence a galaxy to which 
the stellar continuum has been poorly matched. For this we
determined, as a function of the quality of the spectrum, the median
of the rN/sN distribution and corresponding outlier-resistant standard
deviation ($\sigma$). We used the rN/sN distributions in S/sN
bins containing at least 100 objects (see the inset of the central panel
of Fig.~\ref{qa_continuum}), which we determined using the Voronoi
binning scheme\citep{cap03}, and fitted the resulting median and
sigma values (red crosses) with a polynomial function (orange lines).

We then computed the distance of each point from the median line in
Fig.~\ref{qa_continuum}, expressing its distance in units of sigma,
N$\sigma$. Since the rN/sN distributions in each S/sN bin were well
matched by log-normal functions, such N$\sigma$ value readily gave
the probability of a point being a true outlier. For example, N$\sigma$=3 indicated that 
there was a 99.73\% probability that the object was a true outlier 
and only a 0.27\% probability of being the result of random fluctuations.

\subsection{Nebular fit}
\label{ssec:nebfit}

Similar to the case of the stellar continuum, we needed to check the
quality of the fit with the emission lines in order determine possible biases 
in the physical parameters we wanted to measure, such as
the line width or fluxes. Even when the nebular emission was clearly
detected, it was possible that the underlying assumptions in our
modeling approach were not met, due to the presence of a broad
line region or to non-Gaussian line profiles, for example.

In order to assess the quality of the emission-line fit using the
previous approach, it was important to give as much weight as possible
to the deviations that arose only in the spectral regions around the
nebular emission, and to adopt, for the wavelength interval where the
rN/sN ratio was computed, a width that scaled with the width of the
lines.
Indeed, if we were to evaluate the rN/sN ratio over a fixed number of
sampling elements, at a given fit quality the deviations from the fits
of lines of very different widths would contribute unevenly to the
rN/sN budget, thus limiting our ability to spot bad fits in the case
of narrow lines.
For each emission line we thus computed rN/sN around the line within
a passband of width equal to five times the gas velocity dispersion \Sg.
To then assess the quality of an emission-line fit in a given object,
we normalized the distance from the median line in the rN/sN vs. S/sN
diagram using the $\sigma_{\rm rN/sN}$ value (at a given S/sN) that
would be found if all of the rN/sN and S/sN ratios had been measured with
passbands of identical widths corresponding to the value of the \Sg\ 
observed in the object of interest.
Such a $\sigma_{\rm rN/sN}$ was found by simply rescaling
the $\sigma_{\rm rN/sN}$ values obtained using a fixed
bandwidth, corresponding, for instance, to the typical value of
\Sg\ found in the SDSS galaxies ($\sim90$ \kms).  For
a good fit, the amplitude of the fluctuations around the median value
for the rN/sN ratio (at a given data quality or S/sN value) scaled
simply with the square root of the number of sampling elements used in
computing the rN/sN ratio.

There was a second complication in assessing the quality of the
emission-line fit. Even for systems with similar line widths,
evaluating the amplitude of the statistical fluctuations in the rN/sN
ratio at a given data quality was difficult due to the presence of a
larger fraction of formally bad fits compared to the case of the
stellar continuum. Put another way, it was more difficult to
isolate the tail of the outliers when estimating the standard deviation in
the rN/sN histograms for the emission-line fit.
To circumvent this problem, for every line that we wanted to check
in terms of the quality of our fit, we defined a continuum passband in its immediate
vicinity and used the rN/sN vs. S/sN distribution for this passband to
determine the reference rN/sN median and sigma profiles that were
actually used to assess the quality of each fit.  As previously mentioned, 
such profiles were derived by adopting for the continuum passband a single
width, which was set to five times the typical \Sg\ value in the SDSS
galaxies, or approximately 90\kms.

Fig.~\ref{qa_emission} illustrates how this procedure works for the
specific case of the \OIII\ emission line. The left panel shows, for
the same $\sim$46000 galaxies shown in Fig.~\ref{qa_continuum}, the
rN/sN vs. S/sN distribution for the \OIII\ fit, adopting for each
object a bandwidth equal to five times $\sigma_{[\mbox{O{\sc iii}}]}$.
In this panel, we show the positions of three objects with similar
values of the rN/sN ratio around the \OIII\ line, but with different
$\sigma_{[\mbox{O{\sc iii}}]}$ values, as indicated by the size of the
symbols.
%
%\placefigthree
%----------Figure 4
\begin{figure*}
\centering
\includegraphics[width=1\textwidth]{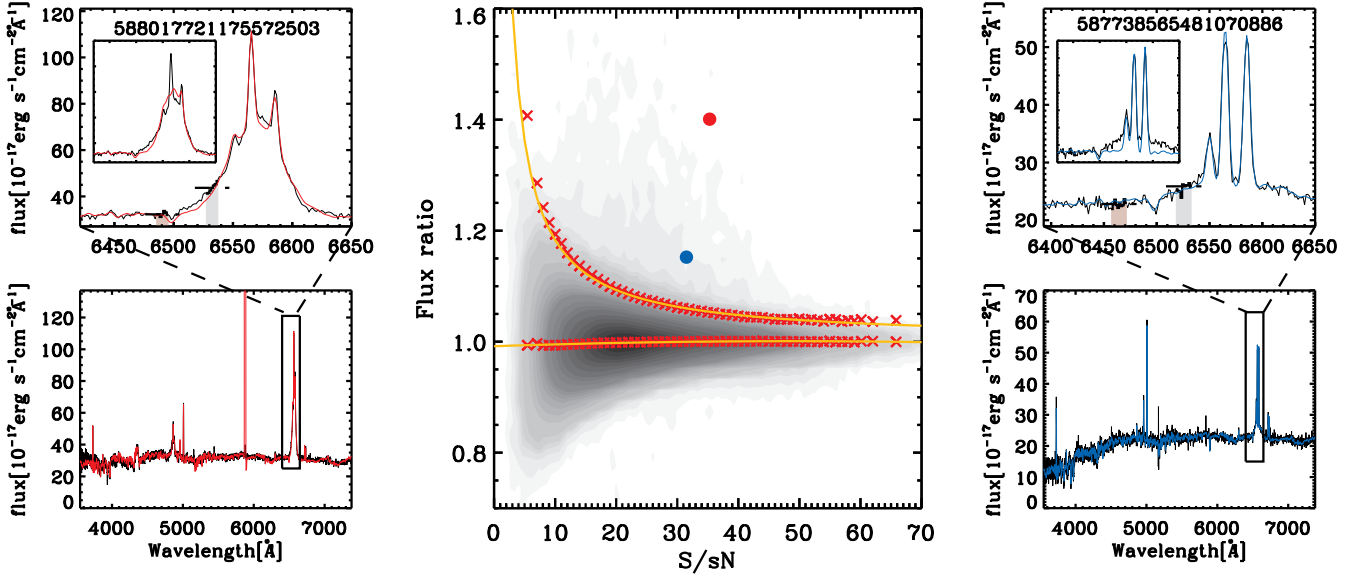}
\caption{Broad line region objects.  The central panel shows how we
  picked out broad line regions from the entire DR7 galaxies. The median
  and 3$\sigma$ distributions are shown with red crosses and orange
  lines, which are shown on left and right panels. Left : The black and red
  lines represent the observed spectrum and fit, respectively. The two shaded
  regions on the top panel are the continuum and blue-side wing of the broadened
  emission region that was used to derive the flux ratio.  The dashed horizontal
  bars represent the mean flux levels for the specifically defined continuum
  and wing. The inserted panel on the top is the fitting result, without the 
  new prescription for the broad line region objects. Once we picked 
  the broad line region objects from the entire DR7 galaxies, we re-fitted
  them to improve the fit. Right : The blue line is our fit. }
 \label{ex_blr}
\end{figure*}

%----------Figure 5
\begin{figure*}
\centering 
\includegraphics[width=0.7\textwidth]{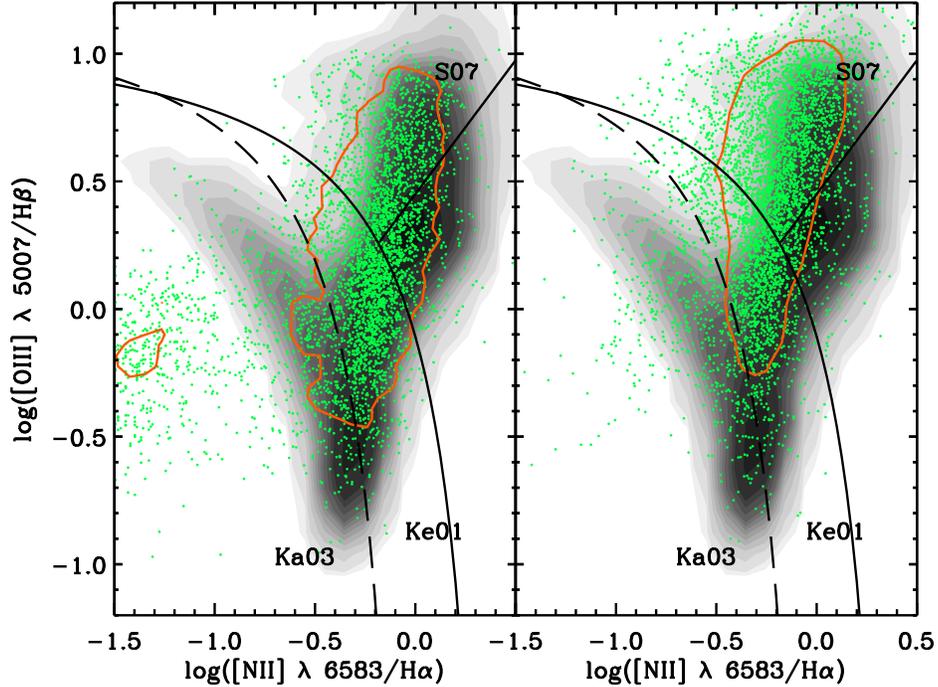}
\caption{BPT diagnostic diagram including broad line regions.
  The objects morphologically classified into early type galaxies by \fracDeV\ 
  and the concentration index were drawn with filled contours. 
  We determined the broad line regions from the ``galaxies'' in the SDSS DR7, which is 
  shown in green dots in both panels. Many of the green dots moved to Seyfert
  regimes at the right panel, by applying a new prescription which masked the 
  entire Balmer absorption series to a properly find continuum. Broad
  line regions, with the exception of widely broadened emission lines, need this new prescription 
  to exactly fit their continuum.}
\label{bpt_blr}
\end{figure*}

If the median and $\sigma_{\rm rN/sN}$ profiles could be drawn in this
panel and used to assess the quality of the \OIII\ fit in these
objects, we would have been led to believe that the goodness of the
emission-line fit degrades as we move from left to right, to higher
S/sN values. This is not quite the case, as illustrated by the central panel.
This panel shows the rN/sN vs. S/sN distribution obtained in the
continuum passband adjacent to the \OIII\ line, centered at
5006.7\AA\ and adopting a width equal to five times 90\kms. 
The dashed and solid black lines show the median rN/sN values and the level 
where the objects are one $\sigma_{\rm rN/sN}$ away from the median,
respectively (using the same method as for the continuum fit).
For lines that are considerably broader than 90\kms\ the statistical
fluctuations around the median ought to be considerably smaller since
a broader passband containing a larger number of sampling elements was adopted. 
Conversely, the quality of the fit for much narrower lines
should be compared against a higher bar.
This is illustrated by the coloured solid lines in the central panel
of Fig.~\ref{qa_emission}, which shows the level where objects with
the same $\sigma_{[\mbox{O{\sc iii}}]}$ of the three considered examples
should lie when found one $\sigma_{\rm rN/sN}$ away from the median
rN/sN line.
When such yardsticks are used to express the distance from
the median in terms of $N_\sigma$ values, although
the object with highest S/sN ratio (green point) lies further away from
the median line, due to its narrower \OIII\ line, the quality of its \gandalf\ fit 
was not much worse than in the case of the galaxy with the worse data (lowest S/sN, light
blue) and was much better than for the object with the broadest
\OIII\ lines (light green).
Using this quality-assessment scheme, the right-hand panel shows 
how the goodness of our emission-line fit can be illustrated
(albeit only for a small number of objects for the sake of clarity), by
colour-coding each point in the rN/sN vs. S/sN diagram for the
\OIII\ line according to the $N_\sigma$ values derived in the rN/sN
vs. S/sN diagram for the continuum band adjacent to \OIII.

Typically, formally bad emission-line fits occur either in the
presence of strong lines where even small deviations from a Gaussian
profiles lead to high rN/sN values, or in the presence of an additional
BLR component.
Non-Gaussian lines can have a double-horned shape. For instance, in the
case of the Balmer lines, tracing star-formation activity confined to a
circumnuclear ring can show blue or red wings, presumably due to out
or inflows of gaseous material, or, can display the triangular
Voigt profile that is typical of strong AGN activity.
Our Gaussian fit could lead to biased measurements for the position of
the lines with blue or red wings, and to an overestimation of the width of the
true thermal component in lines with a Voigt profile. 
Our flux estimates, on the other hand, are not heavily affected by the
presence of non-Gaussians profiles.
This is not the case, however, for objects with an additional BLR
component, in particular when \gandalf\ chooses to match the broad
lines rather than the narrow component of the Balmer spectrum. Since
the BLR fluxes can greatly exceed the flux of the narrow Balmer lines,
when these objects are plotted on standard BPT diagrams, such as the
\NII/\Ha\ vs. \OIII/\Hb\ diagram, they can be easily spotted in
regions with very low values for the \NII/\Ha\ or \OIII/\Hb\ ratio
that are not occupied by other galaxies with well-matched nebular
spectra from AGNs or star-forming regions (refer to the following section).
Conversely, no object with a non-Gaussian profile ends up in similarly
strange regions of the BPT diagrams.

\subsection{Broad Line Region}
\label{ssec:BLRfit}
In the previous two subsections, we argued that the presence of a
broad-line region frequently causes a bad fit to both the emission
lines and the stellar continuum, in particular near the \Ha\ and
\Hb\ lines.
Yet, provided that the presence of a BLR is automatically spotted, the
impact of such a component on the \ppxf\ extraction of the stellar
kinematics and the subsequent \gandalf\ fit to both the stellar
continuum and nebular lines can be easily dealt with. In
fact, one needs only to add a second series of broader Balmer lines in
the \gandalf\ setup to place a broader mask around to Balmer lines
during the \ppxf\ fit and then include the BLR component in the
\gandalf\ fit.

In order to recognize the presence of BLR, we looked for a ``shoulder''
in the continuum region adjacent to the \Ha+\NII\ region, which would
otherwise be relatively flat. For this, we computed the ratio between the
flux levels observed in two spectral regions of equal width on the
blue side of the \Ha+\NII\ blend. The first was placed as close as
possible to the \NIIa\ (avoiding any flux from this line even for the
broadest narrow-line region, $\sim 300$\kms), and the second was located
just beyond the extent of the broadest BLR (centering it on the
blue side of the 6594\AA\ absorption line that is predominantly due to
Calcium and Iron).
Then, similar to our quality-assessment procedures, once such flux
ratio was measured in a given object, we simply checked to determine if it was
exceptionally high compared to the distribution of other flux ratios in
spectra of similar quality.
Fig.~\ref{ex_blr} shows two examples of spectra with BLR components 
and how our procedure isolated them. This diagram juxtaposes the
quality of the spectra in the \Ha+\NII\ region (using the S/sN ratio
in the 6000 -- 6200\AA\ window, \S~4.1) and the flux ratio defined
above.
This figure also shows the \gandalf\ fit according to our
standard prescription and, after adding a BLR component, and
illustrates how our standard \gandalf\ fit does not always fail while
choosing to match the BLR region (as already noticed in \S~4.2).

To assess the efficiency of our criterion in spotting BLR components,
we visually inspected the spectra and the residuals of the standard
\gandalf\ fit for 10,000 randomly chosen objects, labeling those where
we recognized the presence of a BLR. Adopting a 3$\sigma$ cut to
isolate exceptionally high values for the previously defined flux
ratio, our procedure selected 83\% of the BLR that we identified by eye,
while wrongly deeming likely the presence of a BLR in 0.8\% of the
objects where we had found none.
Although these figures are quite encouraging, it must be
acknowledged that only the strongest BLR regions can be identified if
the S/sN is small(i.e., poor-quality spectra).
%\placefigfour
%\placefigfive

%
% Plot EW of BLR (e.g. -300\AA )as a function of S/sN
%
Even then, such BLR components appear in a non-negligible fraction of
the SDSS spectra of nearby galaxies, as overall we found 8,610
objects with a BLR component out or the entire DR7 sample, or almost
1.3\% of the galaxies.

As mentioned in \S~4.2, the impact of a failed \gandalf\ fit due to a
BLR can be crucial when classifying the nebular
emission with standard BPT diagrams. In fact, when the
\gandalf\ algorithm opts to fit the BLR region within our standard
prescription, the object will move to a strange region in the
BPT diagram, below and to the left of the area occupied by the star-bursting
systems. Adding a BLR component in the \gandalf\ fit rectifies this
problem, and usually moves the narrow-line components of these
galaxies back to the regions of the BPT diagram that are occupied by active galactic
nuclei, either Seyfert nuclei or low-ionization nuclear emission
regions (LINERs).
Such a correction was particularly evident when considering a sample of
objects that rarely showed signs of central star formation, such as
early-type galaxies.
Fig.~\ref{bpt_blr} shows an area occupied by the early-type galaxies
in our DR7 sample (as morphologically classified in \S~7.2) in the
standard \NII/\Ha\ vs.  \OIII/\Hb\ diagnostic diagram, as well as the
position of those objects where we spotted the presence of BLR, before
and after adding a BLR component in the \gandalf\ fit, in the left and
right panels, respectively. For the vast majority of the objects with
a BLR, our procedure failed to match the narrow \Ha+\NII\ lines in the
standard \gandalf\ fit and they resided in the lower left side of the BPT
diagram.  If we properly dealt with the BLR component, however, they
generally moved to the region occupied by Seyfert nuclei.

%====================================================================
\section{Database}
\label{sec:database}

Our measurements of the stellar and nebular kinematics, for the
strength of both the emission lines and the stellar absorption
features, for the diffuse and nebular reddening components, as well as
our assessments of the quality of the fit to the stellar continuum
and each of the emission lines, formed a rather complicated database.
To facilitate distribution, we decided to organize our \ppxf, \gandalf\ and
absorption-line index measurements in IDL structures, which allowed us to
group all the emission- and absorption-line quantities. Such
structures were then saved in single FITS files for each of the
objects in our public catalogue.
These files can be queried for single or multiple objects by their
SDSS ID, galaxy coordinate and redshift via a web interface, or can be
downloaded in single archive. For the case of single objects, we also
provide a second FITS file showing the \gandalf\ fit to the object of
interest.

%----------Figure 6
\begin{figure*}
\centering
\includegraphics[width=1\textwidth]{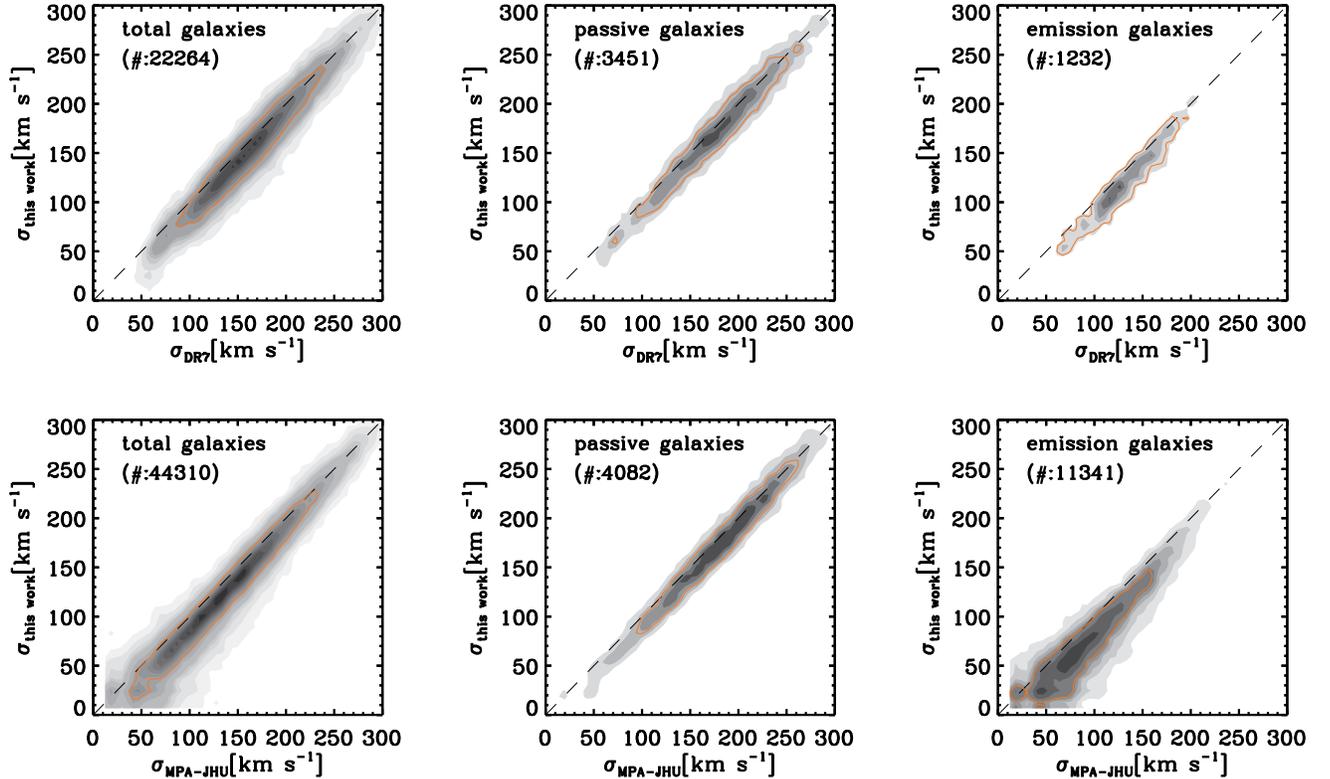}
\caption{Comparison between our values for the central stellar
  velocity dispersion and those computed by the SDSS pipeline (upper
  panels) and those listed in the MPA-JHU catalogue (lower panels).
  Starting from an original sample of randomly drawn 46,000 objects
  from our catalogue, the left panels shows the comparisons with
  matches in the SDSS and MPA-JHU databases, whereas the
  central and left panels shows objects with no nebular emission
  (A/N$<1$ for \OIII, \Ha\ and \NII) or with strong lines (A/N$>4$ for
  the same lines, at the same time), respectively.  In each panel the
  dashed lines show the one-to-one line and the solid lines
  show the 1$\sigma$ contour for the plotted data.}
\label{sigma_comparison}
\end{figure*}

The output parameters in the IDL structures are organized following
the logical order of our fitting procedure. First we found the
\ppxf\ outputs, the stellar redshift $z_{\star}$ and velocity
dispersion $\sigma_{\star}$ (in \kms). These are followed by the
nominal quality of the spectra and of our fit in the continuum, as
quantified by the S/sN and rN/sN ratios introduced in
\S4.1, which relate to the probability that the fit
is biased by some artificial or unaccounted feature, as given
by the N$_{\sigma}$ parameter. Negative N$_{\sigma}$ values occur
when the rN/sN ratio are below the median value, and always indicate a
good fit.
Next are the \gandalf\ emission-line measurements, which we provide
for all of the lines listed in Tab.\ref{tab:emission}, in addition to the
\Hb\ line. For each line we list the observed and de-reddened fluxes
(in $\rm 10^{-17}erg\,s^{-1}\,cm^{-2}\,\AA^{-1}$), the equivalent
width of the line (in negative \AA\ values), the A/N ratio that
relates to how well a line is detected (\S~2.2), and the N$_{\sigma}$
value (as defined in a \S4.2). The strength of the lines and our
ability to match them is then followed by the redshift and width of
the recombination and forbidden lines (labelled as $z_{\rm H\alpha}$,
$\sigma_{\rm H\alpha}$, $z_{\rm [O\,{\sc III}]}$, $\sigma_{\rm
  [O\,{\sc III}]}$) as well as by the value of the interstellar
reddening $A_V$, which includes a second, nebular component
$A_{V,neb}$, when the Balmer decrement is detected (\S~2.2). The
presence of a BLR is also flagged at this point. Although no detection
cut based on the A/N values was applied to the emission-line
measurements, these should be deemed unreliable when little or no
emission is detected.
Last, but not least, are the absorption-line measurements for each of
the indices listed in Tab.~\ref{tab:absorption}, measured in \AA. 
It is most likely that only the objects with SDSS of the best quality (e.g., $\rm S/sN
\ge 30$) would prove useful.

All the previously listed redshift, velocity dispersion, flux,
equivalent width and reddening values are also accompanied by their
corresponding formal uncertainties. In some cases, due to the presence
of sky lines or gaps and artifacts in the SDSS spectra, the values of
the parameters can diverge, in which case they are flagged.

We provide our entire database (line strengths, quality assessments, fitting SED) at 
{\tt{http://gem.yonsei.ac.kr/ossy/}}.

%====================================================================
% MARC
%----------Figure 7
\begin{figure*}
\centering
\includegraphics[width=1\textwidth]{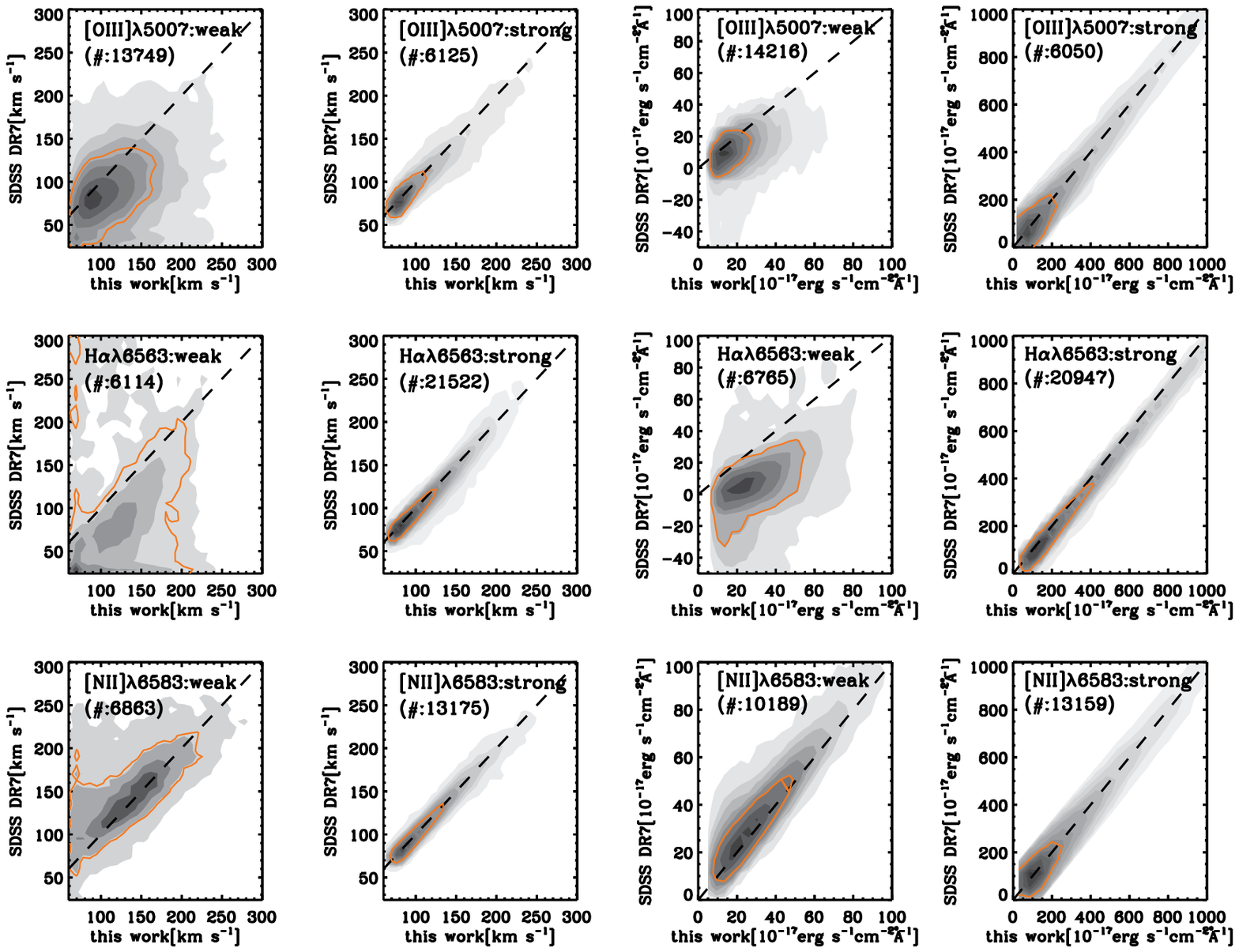}
\caption{Comparison between our values for the emission-line width and
  flux and those from the SDSS pipeline, starting from the same 46,000
  randomly drawn objects used for Fig.~\ref{sigma_comparison}. From
  top to bottom we compared the measurements for the \OIII, \Ha\ and
  \NII\ lines, for the line widths (in the left panels) and
  the line fluxes (in the right ones). The left and right panels are
  further split according to the global strength of the emission (weak : 2$<$A/N$<$5 
  for \OIII, \Ha\ and \NII, strong : A/N$>$10 for the same lines, at the same time). 
  In each panel, the dashed line shows the one-to-one line and the solid contour 
  shows the 1$\sigma$ contour level for the plotted data.}
\label{emission_width_comparison_DR7}
\end{figure*}

%----------Figure 8
\begin{figure*}
\centering \includegraphics[width=1\textwidth]{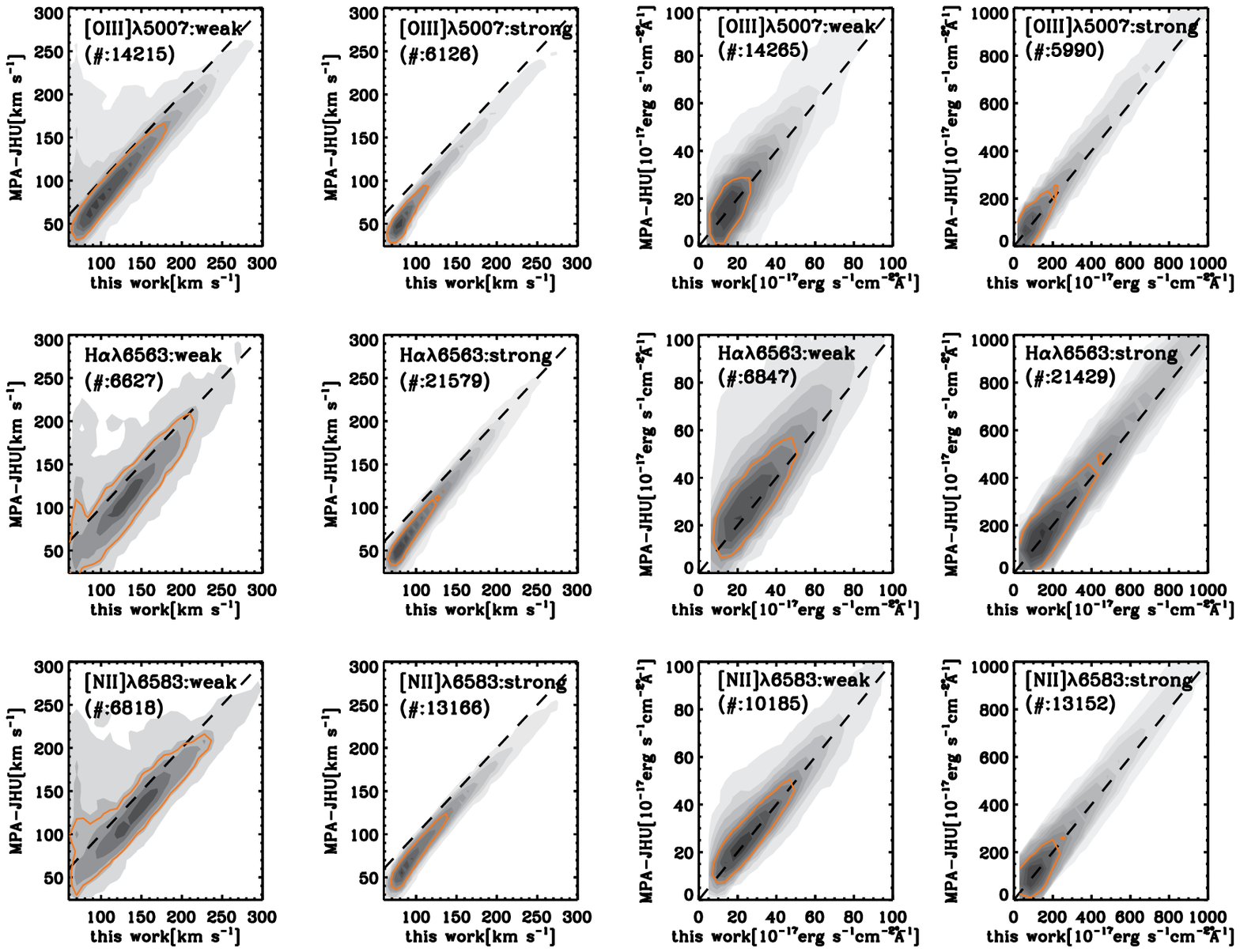}
\caption{Same as Fig.~\ref{emission_width_comparison_DR7}, but showing a comparison 
between the MPA-JHU values and those reported in this work.}
\label{emission_width_comparison_MPA}
\end{figure*}

\section{Comparisons with other public database}
\label{sec:comparison}

There are two existing databases of SDSS DR7 data with which we can compare
our own measurements. The first consists of the SDSS DR7 pipeline
outputs, and the second comes from the MPA-JHU release of DR7 spectral
measurements\footnote{Available at
  \tt{http://www.mpa-garching.mpg.de/SDSS/DR7/}}.
As in previous releases, the SDSS DR7 pipeline does not remove the
nebular emission prior to the absorption-line measurements, so we
expected biases in cases of considerable ionised-gas emission.
On the other hand, by following the method of Tremonti et al. (2004),
the MPA-JHU measurements were based on a procedure that is similar to
ours, whereby the nebular contribution to the spectra and the
corresponding absorption-line infill was carefully treated.
Yet, from the description of the MPA-JHU database, we determined some
technical differences between their procedure and the method adopted here 
which proved important for interpreting possible discrepancies with our measurements.
For instance, we measured the emission-line fluxes while simultaneously matching 
the stellar continuum and the nebular emission with \gandalf, 
whereas the MPA-JHU emission-line measurements were carried
out on the residuals of a previous fit to the stellar continuum,
presumably while masking the regions affected by emission, which we did 
when extracting the stellar kinematics with \ppxf.

% MARC

\subsection{Stellar Kinematics}

With the data in Fig.~\ref{sigma_comparison}, we began by comparing, for the same
subsample of randomly drawn $\sim$46000 objects used in
Fig.~\ref{qa_continuum} and Fig.~\ref{qa_emission}, our
\ppxf\ velocity dispersion $\sigma_\star$ measurements with the
corresponding values from both the SDSS DR7 and MPA-JHU databases.
%
% BEGIN Marc
%
The overall agreement between the various measurements was good,
although our $\sigma_\star$ values appeared to follow more closely the
SDSS DR7 data than the MPA-JHU. This occurred, in particular, at the
low-$\sigma_\star$ regime, which is inhabited mostly by relatively
faint spiral galaxies showing considerable amounts of nebular emission.
In fact, at the very low $\sigma_\star$ end of our sample the presence
of emission caused most of the SDSS DR7 measurements to fail,
and whereas the MPA-JHU catalogue reports $\sigma_\star$ values for
these objects, in most cases their values exceed ours. The large
scatter in this regime can be explained by considering the low S/sN ratio
of the spectra for such faint spirals, but such a bias cannot be
easily explained. It is worthy of note, however, that based on simulations
similar to those presented in \S~2.5, and featuring SDSS spectra for
both very old and young stellar populations, we found the \ppxf\ can
recover $\sigma_\star$ values that scatter evenly around the input
values down to $\sigma_\star =50$\kms\ and S/sN = 10.
%
% END Marc

%\placefigsix
%----------Figure 9
\begin{figure*}
\centering
\includegraphics[width=1\textwidth]{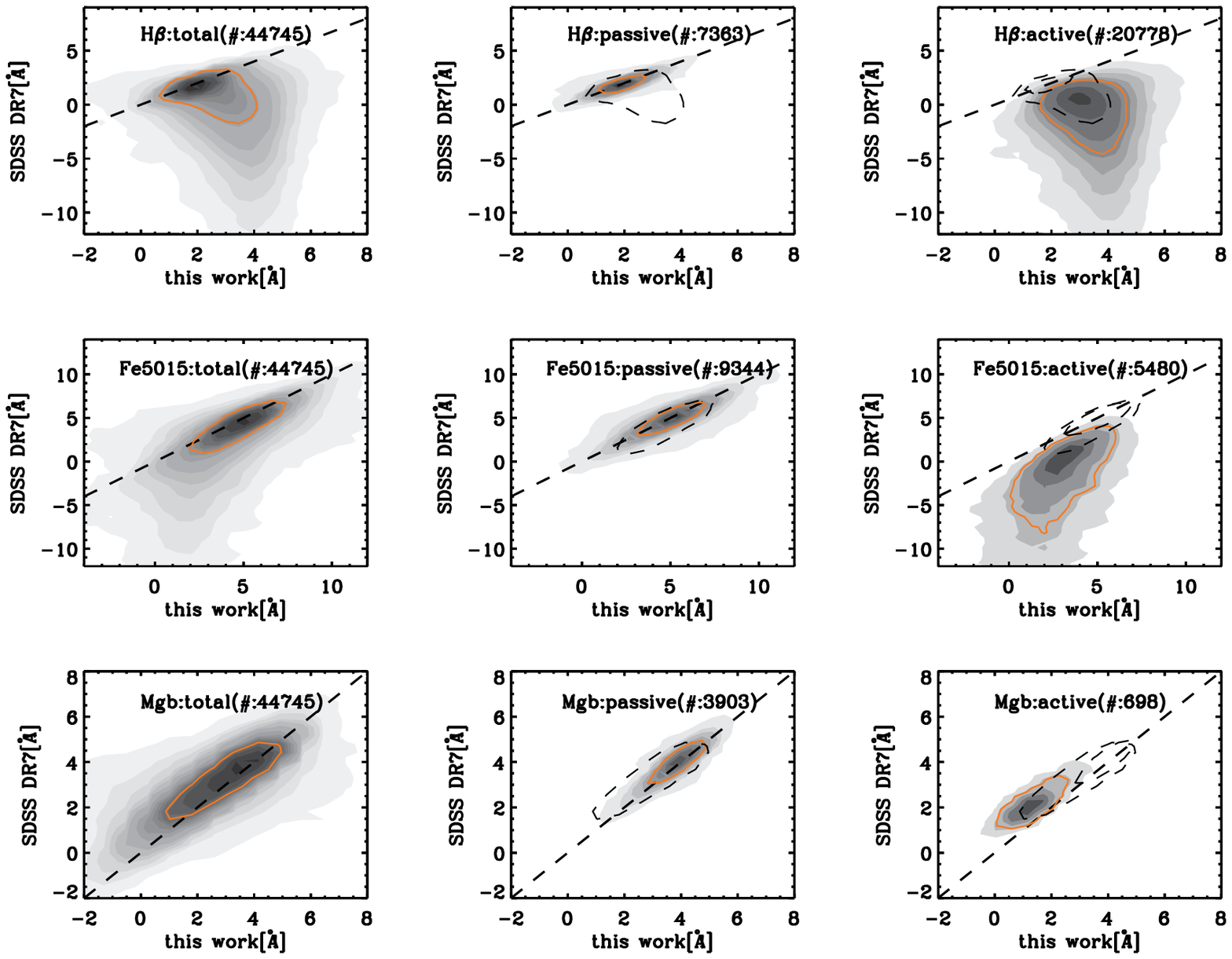}
\caption{Comparison between our values of the \Hb, Fe5015 and Mgb
  indices (top, middle, lower panels) and those from the SDSS
  pipeline, starting from the same 46,000 randomly drawn objects used
  for Fig.~\ref{sigma_comparison}. As in Fig.~\ref{sigma_comparison}, 
  the left panels show all of the objects present in both databases, whereas the
  central and right panels show only the objects with weak nebular
  emission or with strong emission-lines. For the \Hb\ and Fe5015 indices, 
  we defined weak objects as those with A/N values below 1 for the \Ha\ and \OIII\ lines, 
  and active objects as those with A/N$>10$ for those lines. 
  In the case of Mgb, we required intermediate A/N values for \Ha,
  \OIII\ and the \NI\ lines simultaneously to pick objects with weak
  emission. Conversely, we required strong emission in all three
  lines in  order to select objects where the \NI\ contamination to the Mgb index
  was important. In each panel the dashed line shows the
  one-to-one line and the solid contour shows the 1$\sigma$ contour
  level for the plotted data, as in Fig.~\ref{sigma_comparison}, in this figure 
  the dashed lines also show the contours plotted in the previous panels (from left to right).}
\label{DR7_comparison}
\end{figure*}

%----------Figure 10
\begin{figure*}
\centering
\includegraphics[width=1\textwidth]{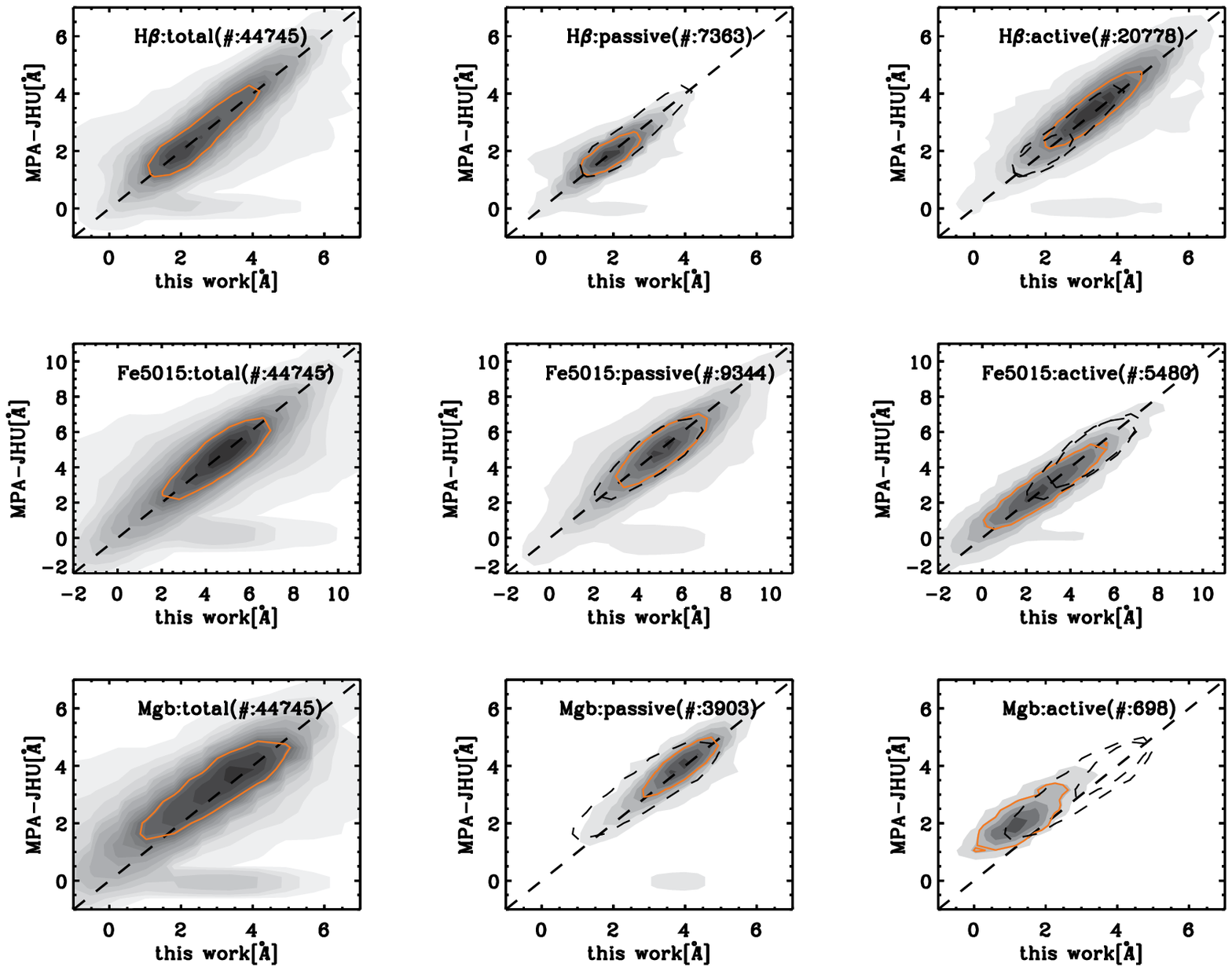}
\caption{Same as Fig.~\ref{DR7_comparison} but showing a comparison 
between the MPA-JHU values and those reported in this work.}
\label{MPA_comparison}
\end{figure*}

\subsection{Nebular Emission Fit}

Continuing with the nebular fit,
Fig.~\ref{emission_width_comparison_DR7} and
Fig.~\ref{emission_width_comparison_MPA} show how our values for the
width and the flux of strong lines such as \OIII, \Ha\ and
\NII\ compare against the same quantities from the DR7 pipeline and
the MPA-JHU catalog, respectively. When comparing our values to the
DR7 and MPA-JHU measurements, we used observed and intrinsic
line-widths, respectively, and in both cases the flux values that were
still affected by reddening due to dust in the SDSS objects. We
further corrected the DR7 fluxes for foreground galactic extinction to
bring them in line with our and the MPA-JHU measurements, and brought 
the MPA-JHU fluxes back to their original values before the
renormalisation for extended sources.

Our line-width and flux measurements agreed fairly well with the DR7
measurements (Fig.~\ref{emission_width_comparison_DR7}), in particular
when the lines were strong and thus the manner by which the stellar
continuum was accounted for should have had only a very limited impact on
the emission-line width and flux estimation based on the Gaussian
fits. Still, even at these regimes our \Ha\ flux values appear
to exceed the DR7 measurements by a few percent.
Furthermore, as we consider weaker lines, the way the \Ha\ flux
measurements compared to each other appeared to be systematically biased,
rather than being simply affected by a larger scatter due to larger
relative errors on the measurements of weak lines in both the DR7 and
our dataset, as seems to be the case for \OIII\ and \NII.
Our flux values became systematically larger than the DR7
measurements as the \Ha\ lines become weaker, which is to be expected
given that we accounted for the presence of the underlying stellar spectrum
and that in this regime the strength of the \Ha\ emission became
comparable to its corresponding stellar absorption feature.

%\placefigseven
%\placefigeight

No such bias as a function of the emission-line strength was observed
when we compared our \Ha\ flux values with the MPA-JHU measurements,
which do account for the stellar continuum
(Fig.~\ref{emission_width_comparison_MPA}), although our intrinsic 
\Ha\ line-width values were larger than the MPA-JHU estimates.
From the limited information provided with the MPA-JHU on-line catalog, 
it is difficult to understand the cause of such small deviations; however, 
in the strong-emission regime where the flux estimates should
be consistent between all three databases, our values were in agreement with the DR7
values, whereas the MPA-JHU measurements were not.

\subsection{Absorption Line Measurements}
\label{sec:abs_measurements}

We conclude this section by presenting Fig.~\ref{DR7_comparison} and
Fig.~\ref{MPA_comparison}, where we compare our line-strength
measurements with the values for the same Lick indices provided by
both the DR7 pipeline and the MPA-JHU catalogue. The DR7 release
webpages specify that the DR7 measurements were made on the original
SDSS spectra, without degrading their resolution to match the
resolution of the Lick/IDS system. Furthermore, a first inspection of the objects
with good quality spectra and that were completely devoid of emission
revealed that this was also the case for the MPA-JHU measurements and
that in both cases no attempt was made to correct the line-strength
measurements for the impact of kinematic broadening. The central
panels of Fig.~\ref{DR7_comparison} and Fig.~\ref{MPA_comparison} show
how, for passive objects, our uncorrected measurements were well-matched with 
both catalogues.

%\placefignine
%\placefigten

When objects with emission are considered, however, the DR7 values
(Fig.~\ref{DR7_comparison}) were heavily biased for the Lick
indices where an emission line could fall either in the index passband
or within the adjacent regions that were used to estimate the level of
the pseudo-continuum above the absorption line region. In fact, during
the DR7 pipeline analysis, the nebular emission was not subtracted prior
to the line-strength measurements. When the emission-line fell within
the index passband, as happened for the \Hb\ and \OIIIb\ lines in
the case of the \Hb\ and Fe5015 indices (top and central panels of
Fig.~\ref{DR7_comparison} and Fig.~\ref{MPA_comparison}), the strength
of the absorption line was under-estimated. When the emission
fell in one of the continuum passbands, as happened for the
\NI\ doublet in the case of the Mgb index (lower panels), the
line-strength was over-estimated, as it was measured against an artificially
enhanced pseudo-continuum level (see also \citealt{ems03}).
On the other hand, the MPA-JHU line-strength measurements tended to
agree with ours in the presence of nebular emission, as they were also 
corrected for line-infill (Fig.~\ref{MPA_comparison}, top and middle panels for \Hb\ and
Fe5015). Yet, the MPA-JHU catalogue does not include several of the
emission lines that are present in our catalogue, such as the
\NI\ doublet, which explains why disagreement between the Mgb
measurements still persists (Fig.~\ref{MPA_comparison} (lower panels)).

% BEGIN MARC
As a final remark, it should be noted that the necessary correction for the impact
of kinematic broadening can be quite severe in the most massive
galaxies with correspondingly high values for the central stellar
velocity dispersion (\citealt{kun04}). This occurs, in particular, for the case
of indices, such as Fe5015 and Mgb, that were typically used to
estimate the metallicity of the degree of $\alpha$-enhancement of
stellar populations, for which not accounting for kinematic broadening
would lead to an underestimation of their values. When we considered that the
MPA-JHU measurements for the Mgb index could have been over-estimated in the
presence of \NI\ emission, we note that using the MPA-JHU values could
lead to over-estimated values of the Mgb/Fe5015 ratio, which is a useful gauge
of $\alpha$-enhancement (\citealt{wor92}).

%----------Figure 11
\begin{figure*}
\centering
\includegraphics[width=1\textwidth]{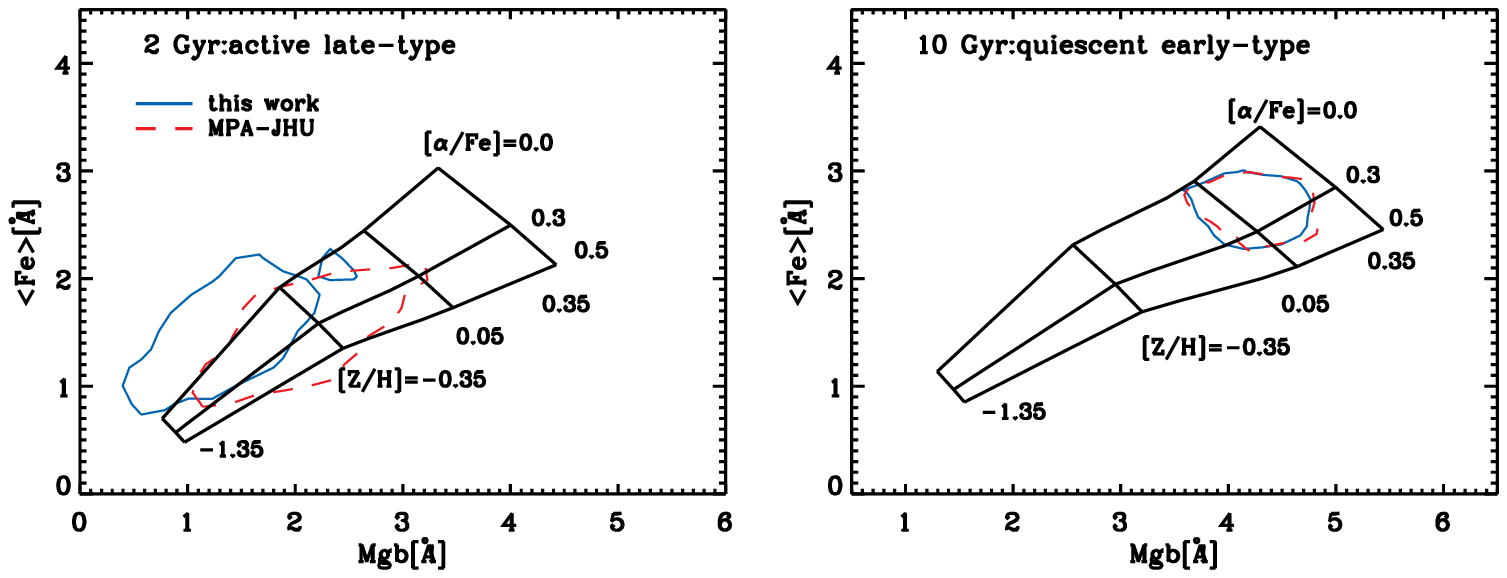}
\caption{Effect of \NI\ emission on the Mgb absorption-line index and
  connected estimates of the $\alpha$-element abundance. The left
  panel shows a comparison of the position of a sample of spiral galaxies with
  strong nebular emission (shown by the 1$\sigma$ contour level of the
  data) in the Mgb vs. $<$Fe$>$ diagram with the model grids of
  \citet{tho03}. The solid and dashed contours indicate the distribution
  of the data from our catalogue and from that of the MPA-JHU
  database. Similarly, the right panel shows the same quantities, but
  for a sample of quiescent early-type galaxies (see \S~\ref{sssec:NI}
  for more detail). The model grids assume a luminosity-weighted
  mean age of 2 Gyr and 10 Gyr for the spirals and early-type galaxies,
  respectively.}
\label{TMB03}
\end{figure*}

\subsubsection{The impact of the \NI\ doublet on element abundance estimates}
\label{sssec:NI}

To give an example of the potential extent of this problem, 
we considered two different subsamples: one comprising spiral galaxies
displaying nebular emission, and a second including only quiescent
early-type galaxies. These objects were morphologically selected as
described in the following section and, after applying our standard
$A/N > 4$ and $A/N < 1$ requirement for strong lines in order to pick
active and quiescent objects (in addition to requiring the detection
or not of \NI\ emission), we assembled 311 and 579 objects for
spirals and early-type galaxies, respectively. Adopting an indicative
luminosity-weighted age of 2 and 10 Gyr for the stellar populations of
these two classes of galaxies, we compared the
position of ours and the MPA-JHU measurements for the Iron and
Magnesium sensitive $<$Fe$>$\footnote{This index is defined in
  \citet{gor90} as $\rm <Fe> = (Fe5270 + Fe5335)/2$} and Mgb indices
to the grids of \citet{tho03} to determine stellar population models of varying
metallicities and overabundance of $\alpha$ elements, such as
Magnesium(Fig.~\ref{TMB03}). For the case of passive objects (right panel), 
the 1-$\sigma$ contours for the distribution of ours and the MPA-JHU
measurements overlapped well and indicated an $\alpha$
enhancement with respect to the solar values of the $[\alpha/{\rm Fe}]$
ratio. On the other hand, for the case of spirals (left panel), 
not accounting for the presence of the \NI\ lines led to an overestimation of the
Mgb/$<$Fe$>$ ratio, and therefore an abundance of $\alpha$ elements
in the central regions of spiral galaxies which, from our database, 
would seem to have solar abundances.

%
% END MARC

%====================================================================
%----------Figure 12
\begin{figure*}
\centering
\includegraphics[width=1\textwidth]{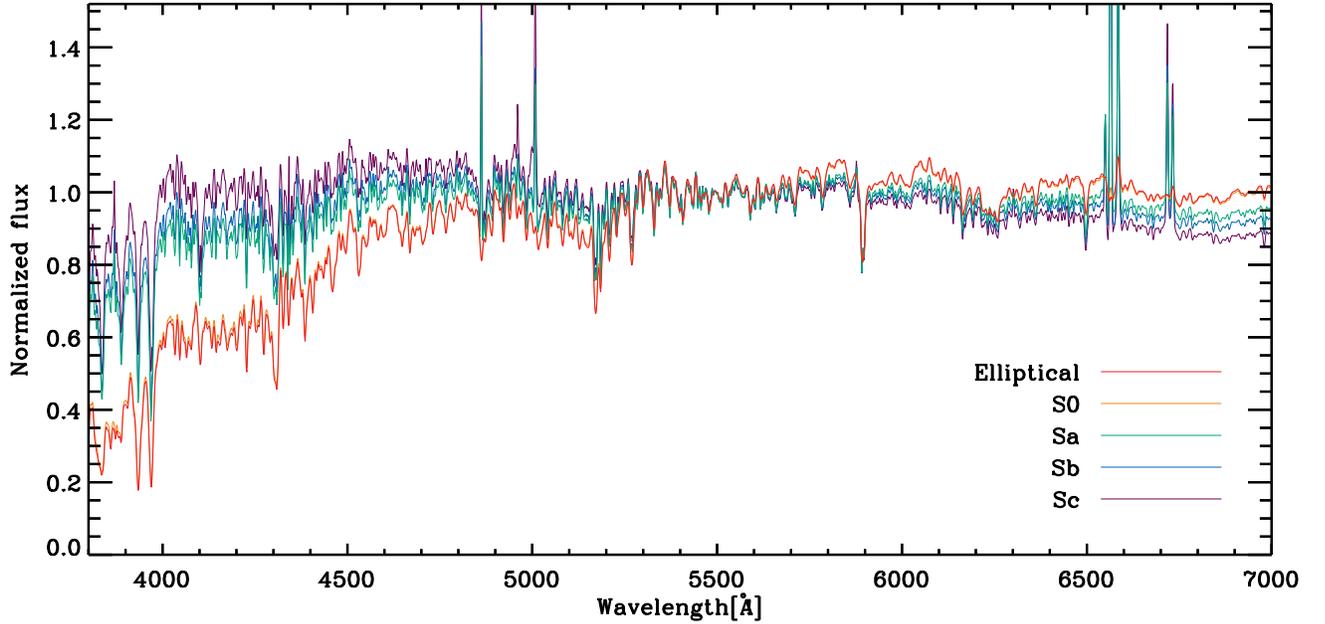}
\caption{Stacked galaxy templates for the various morphological types
  listed in Tab.~\ref{tab:stack} (except edge-on late-type spirals),
  normalized at 5,500 \AA.}
\label{stacked_templates}
\end{figure*}

\section{Galaxy spectral templates}
\label{sec:stack}

Large-scale spectroscopic surveys, such as SDSS, have made it possible
to create spectral templates for galaxies of various kinds and
morphological classes, which can be used as inputs in
SED fitting routines where optical template spectra are needed, 
such as in photometric redshift codes (e.g. \citealt{ste48, bau62, bol00,
  wol04}). Therefore, we present here template spectra of elliptical,
S0, Sa, Sb, Sc, and edge-on late-type galaxies, in the wavelength
range between 3,799\AA\ and 7,000\AA (see Figs. 12, 13 and 14).

We began the construction of our stacked galaxy spectra by restricting
ourselves to a volume-limited parent sample of objects with $M_{r}$
$<$ -17.62 and in the narrow redshift range of 0.020 $<$ z $<$
0.025. This ensured a sufficient quality for the spectra to enable analysis
and stacking, and to ensure that galaxy evolution would not be a concern given than such
a redshift slide covers only $\sim$ 67 Myr.

%----------Table 4
\begin{deluxetable*}{cccccccccc}
\tablewidth{0pt}
\tablecaption{The properties of the sample galaxies used for the SED stacking}
\tablehead{
\colhead{Morphology\tablenotemark{a}} &
\colhead{N\tablenotemark{b}} &
\colhead{$\fracDeV_{g}$\tablenotemark{c}} &
\colhead{$\fracDeV_{r}$\tablenotemark{d}} &
\colhead{$\fracDeV_{i}$\tablenotemark{e}} &
\colhead{$C_{r}$\tablenotemark{f}} &
\colhead{$isoB_{r}$/$isoA_{r}$\tablenotemark{g}} &
\colhead{$isoA_{r}$\tablenotemark{h}} &
\colhead{$R_{\rm eff,r}$\tablenotemark{i}} &
\colhead{$Fraction_{r}$\tablenotemark{j}}
}
\startdata
Elliptical                	& 111 	& 1.0$\pm$0.0 	& 1.0$\pm$0.0 	& 1.0$\pm$0.0 	& 3.3$\pm$0.2 	& 0.8$\pm$0.1  	& 69.1$\pm$19.9  	&   5.9$\pm$1.8 	& 17.1$\pm$4.2      \\
S0                               & 85  	& 1.0$\pm$0.0 	& 1.0$\pm$0.0 	& 1.0$\pm$0.0 	& 3.3$\pm$0.1 	& 0.7$\pm$0.1  	& 68.4$\pm$17.1  	&   5.6$\pm$1.5 	& 19.7$\pm$4.6      \\
edge-on late-type         & 175 	& 0.0$\pm$0.0 	& 0.0$\pm$0.0 	& 0.0$\pm$0.0 	& 2.4$\pm$0.1 	& 0.2$\pm$0.0  	& 61.1$\pm$18.7  	&  23.7$\pm$5.2 	&  8.7$\pm$2.6      \\
Sa                        	& 28  	& 0.0$\pm$0.0 	& 0.0$\pm$0.0 	& 0.0$\pm$0.0 	& 2.0$\pm$0.1 	& 0.6$\pm$0.2  	& 57.6$\pm$13.4  	&  25.2$\pm$4.9 	&  4.6$\pm$1.7      \\
Sb                      		 & 32  	& 0.0$\pm$0.0 	& 0.0$\pm$0.0 	& 0.0$\pm$0.0 	& 2.0$\pm$0.2 	& 0.6$\pm$0.2  	& 60.0$\pm$16.9  	&  26.9$\pm$4.2 	&  4.9$\pm$2.6      \\
Sc                       	 & 21  	& 0.0$\pm$0.0 	& 0.0$\pm$0.0 	& 0.0$\pm$0.0 	& 1.9$\pm$0.2 	& 0.7$\pm$0.2  	& 50.7$\pm$8.2   	&  25.5$\pm$4.3 	&  4.5$\pm$1.2      \\
\enddata
\tablenotetext{a}{All galaxies were visually inspected by eye}
\tablenotetext{b}{Number of stacked galaxy spectra from a complete volume-limited samples}
\tablenotetext{c$\sim$e}{Mean $\fracDeV$ parameters with 1 $\sigma$ standard deviation
in g,r and i bands provided by SDSS pipeline.
In cases of elliptical and S0, all parameters are greater than 0.95 while late-type galaxies have smaller than 0.05.}
\tablenotetext{f}{Mean concentration index($C_{r}$ $\equiv$ $r_{90}$/$r_{50}$)
with 1 $\sigma$ standard deviation in r band.
The value 2.6 used here to classify early-($>$2.6) or late-type($<$2.6) galaxies. }
\tablenotetext{g}{Mean ratio between isophotal major and minor axis with 1 $\sigma$ standard deviation in r band.
All classes have greater than 0.4 of axis ratio except for edge-on late-type($<$0.3).}
\tablenotetext{h}{Mean major axis size with 1 $\sigma$ standard deviation in r band.
All chosen samples are greater than 40 arcsec on its major axis size. }
\tablenotetext{i}{Mean effective radius with 1 $\sigma$ standard deviation in r band. }
\tablenotetext{j}{Mean light fraction(per cent) with 1 $\sigma$ standard deviation in r band.
This parameter shows the light fraction within 1.5 arcsec fiber radius. }
\label{tab:stack}
\end{deluxetable*}

We then applied a set of selection criteria based on SDSS estimates
for the galaxy size, inclination and light concentration in order to
obtain early- and late-type galaxies for which we
could check the morphology by eye. To allow such a
visual inspection, we restricted ourselves to sufficiently
extended and face-on SDSS galaxies, by imposing a lower limit on the
angular size as $isoA_r > 40''$ and the axis ratio as $isoB_r/isoA_r >
0.4$. Furthermore, we initially split early-type galaxies from their
late-type counterparts by looking for objects with concentrated
light profiles that were best matched by a de Vaucouleur profile. For
this we required values for the {\tt FracDeV} parameter, which
measures how well the light profile is matched by a de Vaucouleur law
rather than by an exponential profile, above 0.95 \citep[see
  also][]{sch07a} and values for the concentration index $C_{r}$
($C_{r} = r_{90}$/$r_{50}$ where $r_{90}$ and $r_{50}$ are the radii
containing 90 and 50\% of the light, respectively) above 2.6
\citep[see also][]{shi01,str01}.

Through this pre-selection we obtained 2591 galaxies, which were all
carefully inspected by eye and which in the end yielded 452 objects,
subdivided in E, S0, Sa, Sb and Sc morphologies (as listed in
Tab.~\ref{tab:stack}). This included 175 late-type and edge-on
galaxies as well, which were selected in a similar way, but by imposing
$isoB_r/isoA_r < 0.3$. Our selection criteria is summarized in Tab.~\ref{tab:stack}.

\begin{figure*}
\centering
\includegraphics[width=1\textwidth]{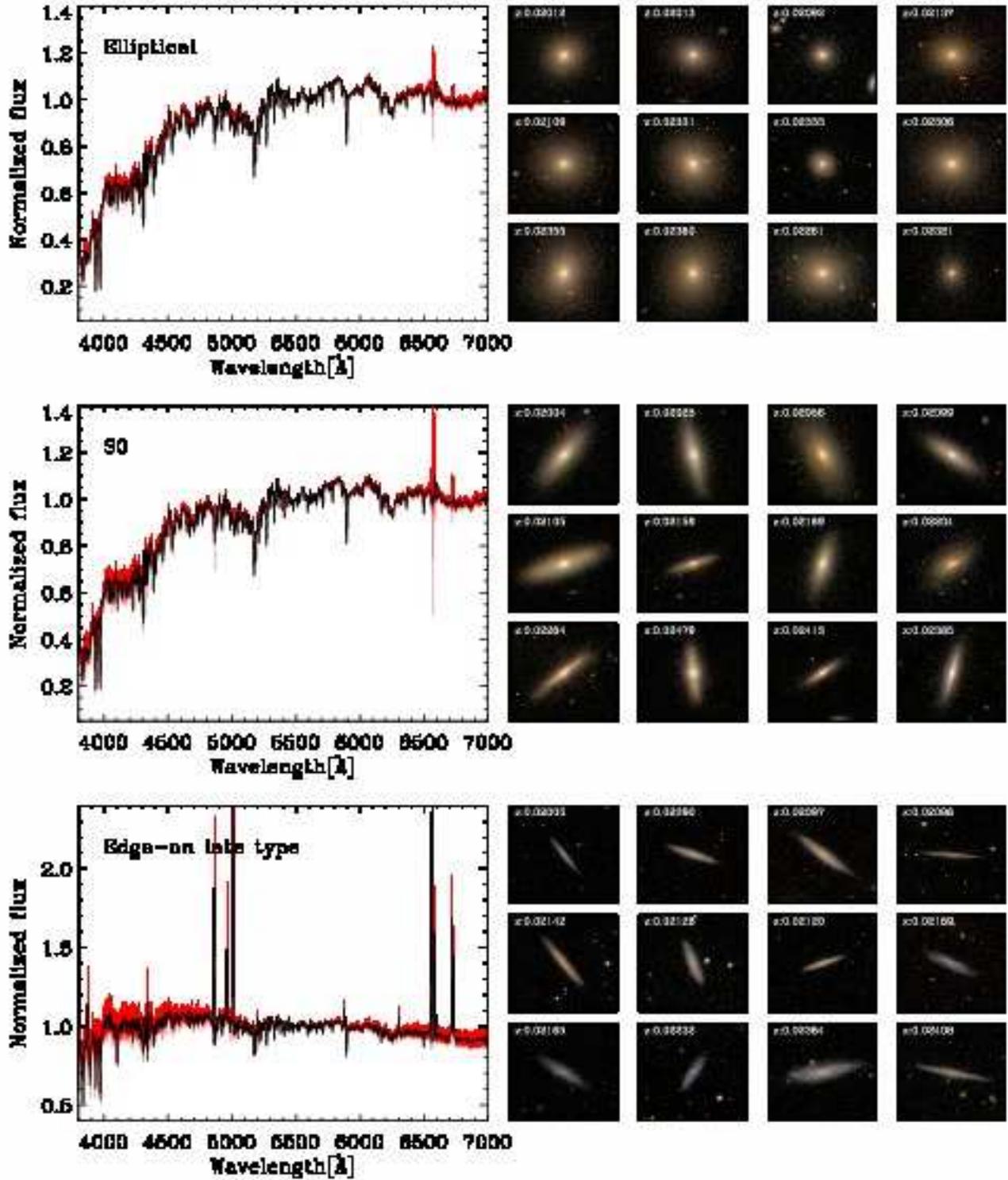}
\caption{Stacked templates and optical images for elliptical,
  lenticular and edge-on late-type galaxy morphologies. The left panel
  includes the 1$\sigma$ scatter estimated while stacking the spectra,
  with the aim of conveying how much difference there was in the shape of
  the spectral energy distribution of the galaxies used in each
  morphological template. As in Fig.~\ref{stacked_templates}, the
  templates were normalized at 5,500 \AA. Each image is
  100.0\arcsec$\times$ 100.0\arcsec\ across, corresponding, on average, 
  to a physical size of 100 kpc.}
\label{stack_early}
\end{figure*}

\begin{figure*}
\centering
\includegraphics[width=1\textwidth]{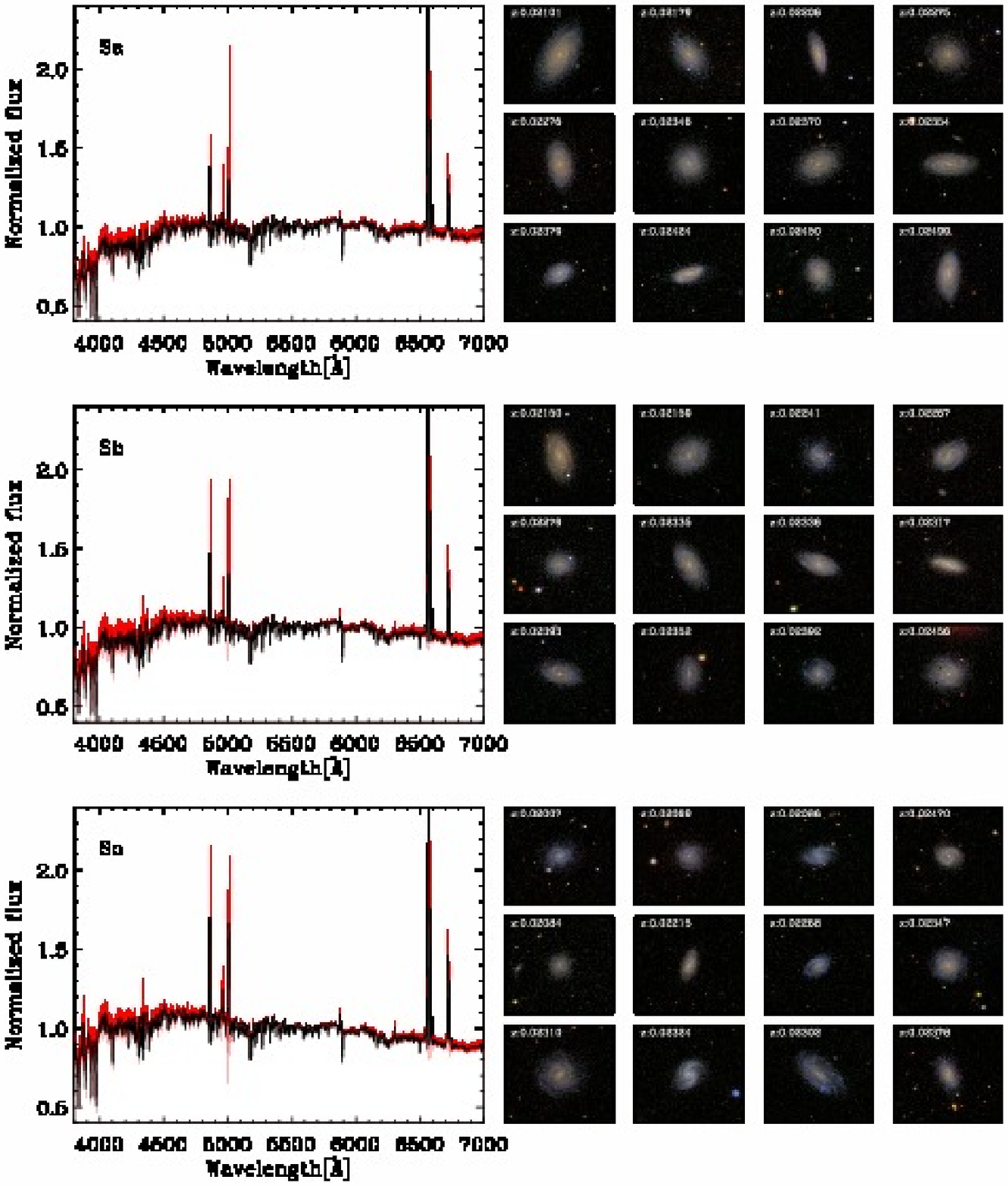}
\caption{Same as Fig.~\ref{stack_early}, but for Sa, Sb and Sc morphologies.}
\label{stack_late}
\end{figure*}

With these robustly classified objects at hand, we proceeded to
combine their SDSS central spectra by stacking the best
\gandalf\ spectral fit for each of them, once each of our models was
brought back to the rest frame and normalized in flux at 5,500\AA. In
this process we also recorded, at each wavelength, the 1$\sigma$ scatter
of the flux values contributed by each of the stacked spectra, in
order to estimate how much the spectral energy distributions of the
galaxies entering our templates differed from each other.
Fig.~\ref{stacked_templates} presents all such spectral templates
(except for the edge-on late-type spirals), which extend from
3,799\AA\ to 7,000\AA, and Figs~\ref{stack_early} and
\ref{stack_late} show each of the spectral templates by itself,
together with SDSS images for the twelve sample galaxies 
in each morphological class that we selected.

As a concluding remark, we note that our stacked spectra encompass
only the central regions of the objects we selected. Specifically, on
average the $3''$-wide SDSS fibers sample 20\% of the light from the E and
S0 galaxies in our sample, and 5\% for the late-type
galaxies. For our E, S0, Sa, Sb, and Sc galaxies, the difference between 
their total magnitudes and their fiber magnitudes is 
0.02, 0.01, 0.09, 0.12 and 0.07 mag, respectively. Obviously, our templates 
are more representative for early-types than for late-type. 
Even for late-types, the systematic trend along the morphological type is 
still visible in our templates, and the systematic offset from the total light properties is 
only of order 10\%. 
%\placefigtwelve
%----------Figure 13

%----------Figure 14

%==============================================================
\section{Summary}
\label{sec:summary}

We presented a new database of absorption and emission-line
measurements for all galaxies within a redshift of 0.2, for which
there are central spectra in 7$^{th}$ SDSS data release.
We modeled the stellar and nebular components of such SDSS spectra, 
adopting publicly available IDL machineries, with the aim of
extracting both reliable and reproduceable measurements for the
stellar velocity dispersion, the strength of various absorption-line
features, and for both the flux and width of the recombination and
forbidden emission lines that were observed in the central region of
our sample galaxies.
In an attempt to improve on existing databases of SDSS spectral
measurements, we included in our fit to the stellar continuum 
both standard stellar population models and empirical templates. 
Our match for the nebular spectrum features an
exhaustive list of both recombination and forbidden lines. 
Foreground dust Galactic extinction is also implicitly treated in our models 
as a simple dust screen component affecting the entire spectrum, 
whereas a second reddening component affects only the nebular spectrum. 

Our absorption-line measurements compare well with the SDSS pipeline
values in the absence of emission. Conversely, the nebular flux
and line-width measurements agree in the strong line regime.
This is consistent with our superior separation of the nebular and
stellar spectra. On the other hand, since The MPA-JHU catalogue is
based on a similarly careful procedure, our measurements also agree 
when considering weak emission, although some emission lines were
omitted in the MPA-JHU catalog. This is the case for the
\NI\ lines in particular, which affect estimates for the $\alpha$-element abundance
of galaxies.

Most importantly, our absorption- and emission-line measurements come
with a quality assessment for the spectral fit that yielded them, in
both the stellar continuum and the nebular emission and across
different wavelength regions of the SDSS spectra.
Such a quality assessment also allows the identification of objects with either
problematic data or with peculiar features that are not accounted for by
the rather standard set of assumptions that enter into our models. For
instance, based on the quality assessment around the \Ha\ and
\NII\ lines, we report here that approximately 1\% of the SDSS spectra
classified as ``galaxies'' by the SDSS pipeline do in fact require
additional broad lines to be matched.

We conclude our work by providing new spectral templates for galaxies of different 
Hubble types, obtained by combining the results of our spectral fits for a subsample 
of 452 morphologically selected objects. As they represent the average spectral 
energy distribution of galaxies of a given Hubble type, these templates should be 
well-suited for photometric redshift estimates based on broad-band fluxes as long as 
the target galaxies appear to have a relaxed morphology. The spectral variations 
with narrower wavelength ranges, due for instance to the impact of a different mass 
on the strength and width of stellar and nebular features, will be the subject of future 
investigations based on larger parent sample for the stacked templates that could be 
selected, for instance, with less stringent morphological or redshift criteria. 

Finally, with the hope of further fostering the discovery potential of our database, 
we also made all of our spectral fits available to the community.

\acknowledgments

We are grateful to the anonymous referee for a number of clarifications that improved
the quality of the manuscript.
We thank Harald Kuntschner for providing us with the
routine that forms the backbone of our absorption-line strength
measurements and to the owner of Caff\`e Caff\`e in Seoul for his kind
hospitality during much of the writing of this paper.
We acknowledge the tireless effort of the SDSS team. The SDSS database
enables countless new investigations on galaxies.  It is our wish to
enhance its usefulness by providing our measurements. 
SKY thanks Richard Kron, Douglas Tucker, Don York and the SDSS team 
at University of Chicago and the Fermilab for constructive feedback and hospitality
during my visit to Chicago.
SKY acknowledges the support by the National Research Foundation of Korea
through the Doyak grant (No. 20090078756), Hakjin grant (KRF-C00156) and
the SRC grant to the Center for Galaxy Evolution Research.
MS acknowledges support from his
STFC Advanced Fellowship (ST/F009186/1) whereas KS acknowledges
support by NASA through an Einstein Postdoctoral Fellowship
(PF9-00069), issued by the Chandra X-ray Observatory Center, which is
operated by the Smithsonian Astrophysical Observatory for and on
behalf of NASA under contract NAS8-03060.

\end{document}